\documentclass[10pt,twocolumn,twoside]{IEEEtran}
\usepackage{algorithmic}
\usepackage{array}
\usepackage{textcomp}
\usepackage{stfloats}
\usepackage{url}
\usepackage{verbatim}
\usepackage{graphicx}
\def\BibTeX{{\rm B\kern-.05em{\sc i\kern-.025em b}\kern-.08em
    T\kern-.1667em\lower.7ex\hbox{E}\kern-.125emX}}
\usepackage{balance}
\usepackage{amsmath,amssymb,amsfonts}
\usepackage{amsthm}
\usepackage{graphicx}
\usepackage[dvipsnames]{xcolor}
\usepackage{mathtools}

\usepackage{hyperref}
\hypersetup{
     colorlinks = true,
     linkcolor = blue,
     filecolor = magenta,      
     urlcolor = cyan,
     citecolor = Blue,
     anchorcolor = green
}

\usepackage{subcaption}
\usepackage{enumitem}
\newtheorem{Definition}{Definition}
\newtheorem{Theorem}{Theorem}

\newtheorem{Lemma}{Lemma}
\newtheorem{Problem}{Problem}

\newtheorem{Remark}{Remark}
\newtheorem{Assumption}{Assumption}

\newcommand{\mc}[1]{\mathcal{#1}}
\newcommand{\mb}[1]{\mathbb{#1}}

\newcommand{\m}[1]{\mathbf{#1}}
\newcommand{\bm}[1]{\boldsymbol{#1}}

\begin{document}
\title{Bearing-Constrained Leader-Follower Formation of Single-Integrators with Disturbance Rejection: Adaptive Variable-Structure Approaches}
\author{Thanh Truong Nguyen, Dung Van Vu,
Tuynh Van Pham, Minh Hoang Trinh,
\thanks{T. T. Nguyen is with Bosch Global Software Technologies, Hanoi, Vietnam. E-mail: \texttt{thanhxetang69@gmail.com}} 
\thanks{D. V. Vu is with Unmanned Aerial Vehicle Center, Viettel High Technology Industries Corporation, Hanoi, Vietnam. E-mail: \texttt{vuvandung.bkhn@gmail.com}}
\thanks{T. V. Pham is with Department of Automation Engineering, School of Electrical and Electronic Engineering, Hanoi University of Science and Technology (HUST), Hanoi, Vietnam. E-mail: \texttt{tuynh.phamvan@hust.edu.vn}}
\thanks{M. H. Trinh is with AI Department, FPT University, Quy Nhon City, Binh Dinh, Vietnam. Corresponding author. E-mail: \texttt{minhtrinh@ieee.org}}}

\markboth{Journal of \LaTeX\ Class Files,~Vol.~18, No.~9, September~2020}%
{How to Use the IEEEtran \LaTeX \ Templates}

\maketitle

\begin{abstract}
This paper studies the problem of stabilizing a leader-follower formation specified by a set of bearing constraints while disturbed by some unknown uniformly bounded disturbances. A set of leaders are positioned at their desired positions while each follower is modeled by a single integrator with an additive time-varying disturbance. Adaptive variable-structure control laws using displacements or only bearing vectors are applied to stabilize the desired formation. Thanks to the adaptive mechanisms, the proposed control laws require neither information on the bearing Laplacian nor the directions and upper bounds of the disturbances.
It is further proved that when the leaders are moving with the same bounded uniformly continuous velocity, the moving target formation can  be achieved under the proposed control laws. Simulation results are provided to support the stability analysis.
\end{abstract}


\section{Introduction}
\label{sec:1}
Recent years have been a booming research period for bearing-based formation control  \cite{zhao2015bearing,trinh2018bearing}, a research topic inspired from the observation that animals can self-localize, navigate, and perform formation-type collective behaviors using their vision power. Research from diverse fields suggests that that fairly simple vision-based guidance rules in animals can unfold sophisticated formation-type phenomena \cite{Loizou2007biologically}. From an engineering perspective, there have been ongoing attempts to understand and realize these displayed formations. Considered as the eye of an autonomous agent (UAV, AGV), the camera provides the bearing vectors (directional information) from the agent to some neighboring agents. In addition to providing an alternative solution for other formation approaches (position-, displacement-, distance-based formation control, etc.) \cite{oh2015survey}, bearing-only control laws are preferred since they reduce the number of sensors used by each agent, cut down on deployment cost, and do not transmit any signals \cite{Ye2017bearing}. In addition, research results from bearing-constrained formation control are applicable to the dual problem - bearing-based localization in wireless sensor networks \cite{Zhao2016aut}.

The theoretical basis of bearing-constrained formation control in $d$-dimensional space ($d\ge2$) was developed in \cite{zhao2015bearing,Zhao2016aut,Zhao2017laman}. Several initial studies on the bearing/directional rigidity theory in two- or three-dimensional space can be found in \cite{Eren2006using,Bishop2014,Eric2014,Tron2015rigid}. As robustness is an importance issue of any multiagent system, consensus and formation control under disturbances were studied by \cite{Cao2011TAC,Hu2019distributed,Chen2022maneuvering,Vu2021ICCAIS}. 
Although disturbances can be actively included for additional objectives such as escaping from an undesired unstable formation \cite{trinh2018bearing}, or formation maneuver \cite{DeMarina2016distributed}, the presence of unknown disturbances usually makes the target formation unachievable or causes unexpected formation motions. Robust bearing-constrained formation acquisition/tracking have recently been proposed in the literature. The works \cite{Li2018bearing,Li2020tcyb,zhao2019bearing,zhao2020bearing,Li2021bearing} assumed the leaders' velocity and the disturbances are constant, or their upper bounds are known by the agents, or the rate of bearings is available. The work \cite{Garanayak2023bearing} proposed an elevation-based bearing-only formation control with disturbance rejection for single- and double-integrators. However, the method in \cite{Garanayak2023bearing} is only effective for minimally rigid formations, and for double integrators, velocity measurements are required. The authors in \cite{Wu2024distributed} studied bearing-only formation control with fault tolerance and time delays. Actuator faults were modeled as a disturbance of unknown control direction, which can be compensated by a control action with an appropriate control gain. The authors in \cite{Bae2022ijrnc} proposed a robust adaptive design method to attenuate the effects of the disturbances to satisfy a specific performance requirement. The authors in \cite{Trinh2021LCSS} considered the bearing-only acyclic formation tracking problem with unknown leaders' velocity using two time-varying gains. Formation maneuver via bearing-only estimation for time-varying leader-follower formations was also proposed in \cite{Huang2021bearing,Su2022bearing,Tang2022localization}. A moving target formation was cooperatively estimated from the measured bearing vectors, and each follower controls its position to track its estimated target position.

This paper focuses on the bearing-based leader-follower formation control problems with single-integrator modeled agents perturbed by unknown and bounded uniformly continuous disturbances. By bearing-based, we assume that the geometric constraints that define the target formation are given as a set of bearing vectors. There are several leaders in the formation, whose positions already satisfy a subset of bearing constraints. The remaining agents, called followers, can measure either (i) the relative positions (displacement-based control) or (ii) the bearing vector (bearing-only control) to their neighbors. The interaction topology between agents is not restricted into an acyclic graph, but is applicable to any infinitesimal bearing rigid formation having at least two leaders. 

Unlike \cite{Tran2018TCNS}, where a disturbance-free finite-time bearing-only formation control was studied or a small adaptive perturbation was purposely included to globally stabilize the target formation in finite time, the disturbances in this work represent unmodeled dynamics or the effects of the environment. The problem is firstly solved under the assumption that the agents can measure the relative displacements. The solution for relative-displacement provides hints for the more challenging task of stabilizing the desired formation when agents can only sense the bearing vectors. Intuitively, since no information on the distances is available, in order to suppress the disturbances with unknown magnitude, the gain of the bearing-only control law should be increased whenever all bearing constraints are not satisfied. This intuition is mathematically realized by adaptive variable-structure control (also known as adaptive sliding mode control), which can provide fast convergence and robustness with regard to disturbances \cite{Oliveira2016adaptive,Roy2020adaptive}. The main novelty of the proposed control laws is providing a distributed adaptive mechanism that alters the magnitude of the control law with regard to the errors between the desired and the measured bearing constraints. In this way, the control input eventually approximates the magnitude of the disturbance, rejects the disturbance and stabilizes the target formation without requiring any inter-agent communication, a priori information on the upper bound of the disturbance, or the formation's rigidity index.\footnote{Specifically, the smallest eigenvalue of the grounded bearing Laplacian is not needed for stabilizing the formation under unknown disturbances.} Modifications of the  control laws are proposed to alter the adaptive gains based on the disturbance's magnitude or to stabilize the target formation in case the upper bound of the unknown disturbance is a polynomial of the formation's error. Moreover, when the leaders move with the same bounded uniformly continuous velocity, their motions can be considered as disturbances to the bearing errors dynamics of the followers. Thus, the proposed adaptive control laws can also be applied to stabilize a time-varying target formation. To sum up, for formations of single integrators, the proposed control laws provide a unified solution to two problems: leader-follower formation control with unknown disturbances and formation tracking with unknown leaders' velocity.

The rest of this paper is organized as follows. Section~\ref{sec:2} presents theoretical background on bearing rigidity theory and formulates the problems. Sections~\ref{sec:3} and \ref{sec:4} propose  formation control/tracking laws using only displacements and/or only bearing vectors, respectively. Section~\ref{sec:5} provides numerical simulations. Lastly, section~\ref{sec:6} concludes the paper.

\emph{Notations.} In this paper, the set of real numbers is denoted by $\mb{R}$. Scalars are denoted by small letters, and vectors (matrices) are denoted by bold-font small (capital) letters. For a matrix $\m{A}$, we use $\text{ker}(\m{A})$, $\text{im}(\m{A})$ to denote the kernel and the image of $\m{A}$, and rank$(\m{A})$ denotes the rank of $\m{A}$. The 2-norm and 1-norm of a vector $\m{a} = [a_1,\ldots,a_n]^\top \in \mb{R}^d$ are respectively denoted as $\|\m{a}\| = \sqrt{\sum_{i=1}^d a_i^2}$ and $\|\m{a}\|_1 = {\sum_{i=1}^d |a_i|}$. The $d\times d$ identity matrix is denoted by $\m{I}_d$, $\m{0}_{a\times b}$ denote the $a\times b$ zero matrix, and $\m{0}_d$ denotes the $d$-dimensional zero vector.

\section{Problem statement}
\label{sec:2}
\subsection{Bearing rigidity theory}
Consider a set of $n$ points in $d$-dimensional space ($n\ge 2,~d\ge2$). The points are positioned at $\m{p}_i\in \mathbb{R}^d$, with $\m{p}_i \neq \m{p}_j,~\forall i \neq j,~ i, j \le n$. A framework (or a formation) in the $d$-dimensional space  ($\mc{G},\m{p}$) is specified by an undirected graph $\mc{G}=(\mc{V},\mc{E})$ (where $\mc{V}$ is the vertex set of $|\mc{V}|=n$ vertices and $\mc{E}$ is the edge set of $|\mc{E}|=m$ edges {without self-loops}) and a configuration $\m{p} = [\m{p}_1^\top,\ldots, \m{p}_n^\top]^\top \in \mathbb{R}^{dn}$. The neighbor set of a vertex $i\in \mc{V}$ is defined by $\mc{N}_i = \{j\in \mc{V}|~(i,j) \in \mc{E}\}$. The graph $\mc{G}$ is connected if for any two vertices $i,j\in \mc{V}$, we can find a sequence of vertices connected by edges in $\mc{E}$, which starts from $i$ and ends at $j$.

Let the edges in $\mc{E}$ be indexed as $e_1,\ldots,e_m$. For each edge $e_k = (i,j) \in \mc{E},~k=1,\ldots,|\mc{E}|=m$, the bearing vector pointing from $\m{p}_i$ to $\m{p}_j$ is defined by 
$\m{g}_{ij} \equiv \m{g}_k = \frac{\m{z}_{ij}}{\Vert \m{z}_{ij}\Vert}$, with $\m{z}_{ij} \equiv \m{z}_k= \m{p}_j-\m{p}_i$ is the displacement vector between $i$ and $j$. It is not hard to check that $\|\m{g}_{ij}\| = 1$, where $\|\cdot \|$ denotes the 2-norm. An edge $e_k =(i,j)$ is oriented if we specify $i$ and $j$ as the start and the end vertices of $e_k$, respectively. According to an arbitrarily indexing and orienting of edges in $\mc{E}$, we can define a corresponding incidence matrix $\m{H}=[h_{ki}] \in \mb{R}^{m \times n}$, where $h_{ki}=- 1$ if $i$ is the start vertex of $e_k$, $h_{ki}=+1$ if $i$ is the end vertex of $e_k$, and $h_{ki}=0$, otherwise. Then, we can define the stacked displacement vector $\m{z}=[\ldots,\m{z}_{ij}^\top,\ldots]^\top= [\m{z}_1^\top,\ldots,\m{z}_m^\top]^\top = \bar{\m{H}}\m{p}$, where $\bar{\m{H}}= \m{H}\otimes \m{I}_d$.

For each bearing vector $\m{g}_{ij} \in \mb{R}^d$, we define a corresponding projection matrix $\m{P}_{\m{g}_{ij}} = \m{I}_d-\m{g}_{ij} \m{g}_{ij}^\top$. The projection matrix $\m{P}_{\m{g}_{ij}}$ is symmetric positive semidefinite, with a unique zero eigenvalue and $d-1$ unity eigenvalues. Moreover, the kernel of $\m{P}_{\m{g}_{ij}}$ is spanned by $\m{g}_{ij}$, i.e., ker$(\m{P}_{\m{g}_{ij}})=$im$(\m{g}_{ij})$.

\begin{figure}
\centering
\begin{subfloat}[]{
\includegraphics[width=0.19\textwidth]{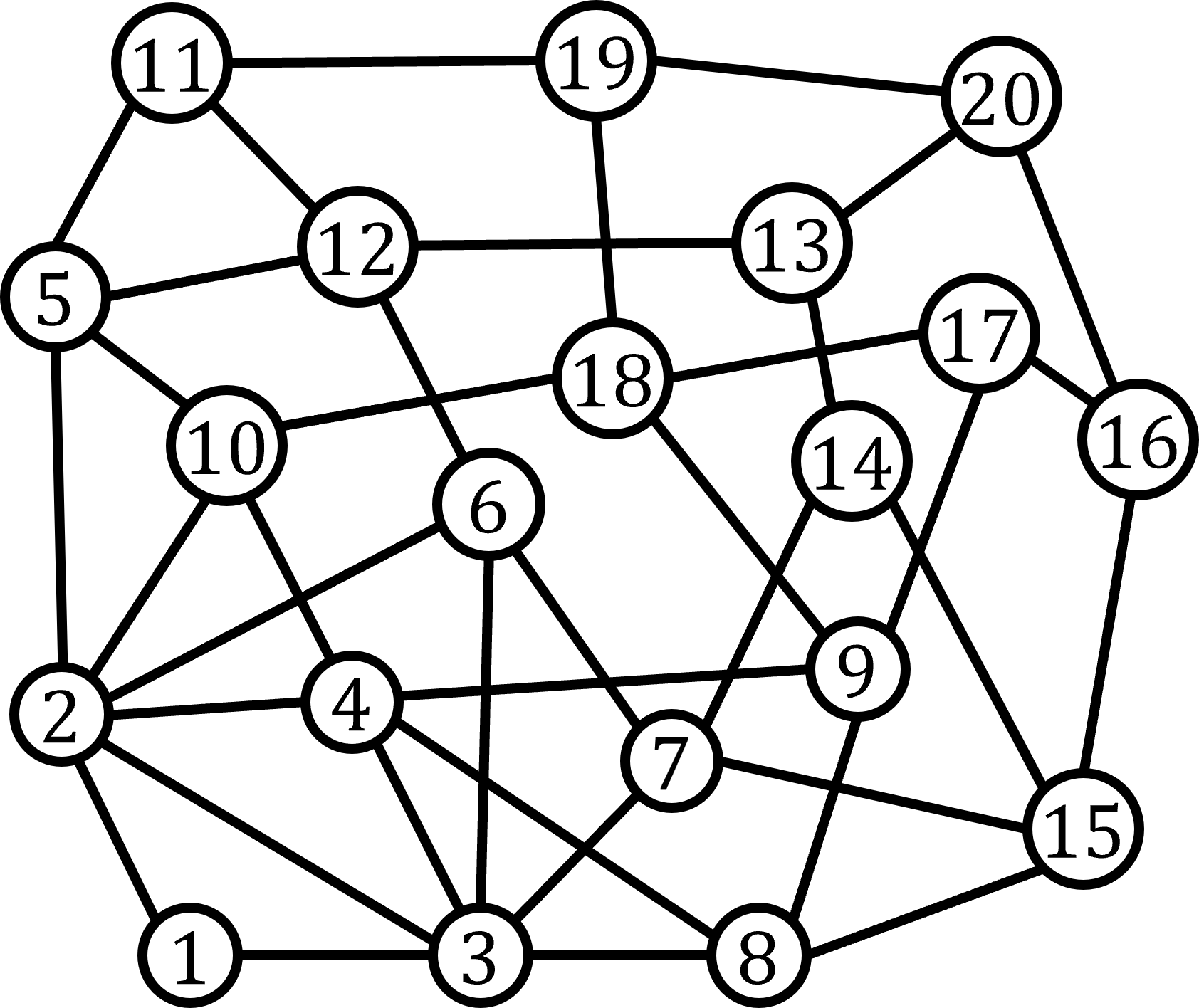}
\label{fig:graphG}}
\end{subfloat}
\begin{subfloat}[]{
\includegraphics[width=0.26\textwidth]{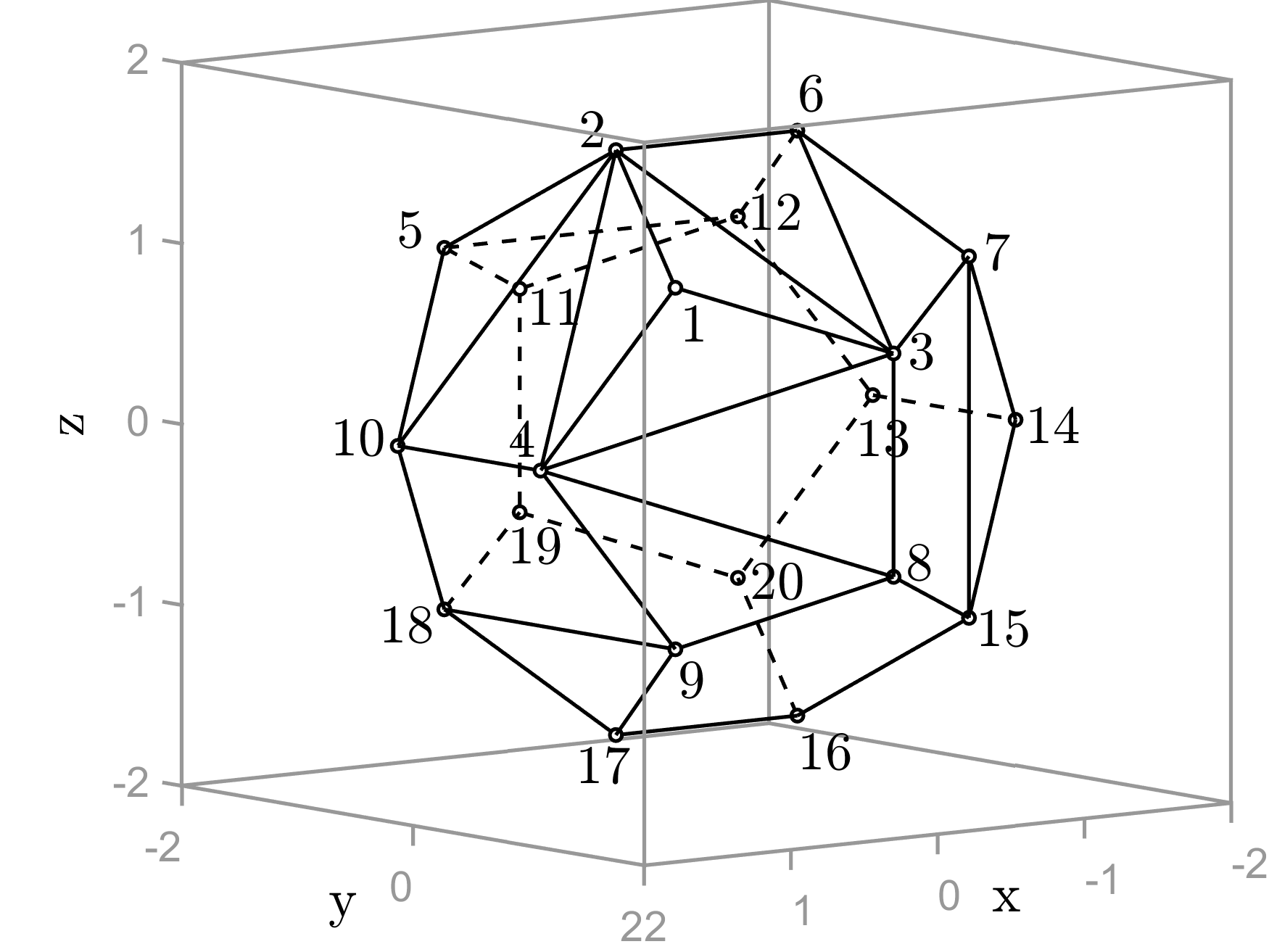}
\label{fig:configuration}}
\end{subfloat}
\caption{An infinitesimally bearing rigid framework $(\mc{G},\m{p}^*)$ in $\mb{R}^3$. (a) the graph $\mc{G}$; (b) a desired configuration $\m{p}^*$ where $\m{p}_i^*, i=1,\ldots, 20,$ are located at the vertices of a dodecahedron.\label{fig:desiredFormation}}
\end{figure}

Two formations ($\mc{G}, \m{p}$) and ($\mc{G}, \m{p}'$) are bearing equivalent if and only if: $\m{P}_ {\m{g}_{ij}} (\m{p}_j'-\m{p}_i') = \m{0}_d, ~\forall (i, j) \in \mc{E}$. They are bearing congruent if and only if 
$\m{P}_{\m{g}_{ij}} (\m{p}_j'-\m{p}_i') = \m{0}_d,~ \forall i, j \in \mc{V}, i \neq j$. A formation ($\mc{G}, \m{p}$) is called globally bearing rigid if any formation having the same bearing constraints with $(\mc{G}, \m{p})$ is bearing congruent with $(\mc{G},\m{p})$. 
Let $\m{g} = [\m{g}_1^\top,\ldots,\m{g}_m^\top]^\top \in \mathbb{R}^{dm}$, the bearing rigidity matrix is defined by
\begin{align*}
    \m{R}_b(\m{p})= \frac{\partial \m{g}}{\partial \m{p}}=\text{blkdiag}\left(\frac{\m{P}_{\m{g}_k}}{\Vert \m{z}_k\Vert}\right)\Bar{\m{H}}\in \mathbb{R}^{dm\times dn}.
\end{align*}
A formation is infinitesimally bearing rigid in $\mathbb{R}^d$ if and only if $\text{rank}(\m{R}_b) = dn-d-1$, this means $\text{ker}(\m{R}_b) = \text{im}([\m{1}_n\otimes \m{I}_d, \m{p}-\m{1}_n\otimes \bar{\m{p}}])$, where $\bar{\m{p}}=\frac{1}{n}(\m{1}^\top_n\otimes \m{I}_d)\m{p} = \frac{1}{n}\sum_{i=1}^n \m{p}_i$ is the formation's centroid. An example of infinitesimally bearing rigid framework is shown in Fig.~\ref{fig:desiredFormation}.

In bearing-based formation, we usually use an augmented bearing rigidity matrix $\tilde{\m{R}}_b=\text{blkdiag}(\Vert \m{z}_k\Vert \otimes \m{I}_d)\m{R}_b=\text{blkdiag}(\m{P}_{\m{g}_k})\Bar{\m{H}}$, which has the same rank as well as the same kernel as $\m{R}_b$ but does not contain information of the relative distances between the agents $\|\m{z}_k\|$. Further, we define the bearing Laplacian ${\m{L}}_b(\m{p}) = \tilde{\m{R}}_b^\top \tilde{\m{R}}_b$ which is symmetric and positive semidefinite. For an  infinitesimally rigid formation, $\m{L}_b$ has exactly $d+1$ zero eigenvalues and ker$({\m{L}}_b)=$ker$(\m{R}_b)$.

\subsection{Problem formulation}
Consider a system consisting of $n$ autonomous agents in the $d$-dimensional space ($d\geq 2$), of which the positions are given by $\m{p}_1,\ldots,\m{p}_n \in \mb{R}^d$. We assume that there exist $2\leq l < n$ stationary leader agents in the formation and the remaining $f=n-l$ agents are followers. 

Defining the vectors $\m{p}^{\rm L}=[\m{p}_1^\top,\ldots,\m{p}_l^\top]^\top \in \mb{R}^{dl}$ and $\m{p}^{\rm F}=[\m{p}_{l+1}^\top,\ldots,\m{p}_{n}^\top]^\top \in \mb{R}^{df}$. When the leaders are stationary, we can write $\dot{\m{p}}^{\rm L}=\m{0}_{dl}$. 

The follower agents are modeled by single integrators in the $d$-dimensional space with the disturbances:
\begin{align} \label{eq:agent_model}
    \dot{\m{p}}_i(t) = \m{u}_i(t) + \m{d}_i(t),\, i = l+1, \ldots, n,
\end{align}
where $\m{p}_i$ and $\m{d}_i$ denote the position and the disturbance of agent $i$, respectively.

The desired formation $(\mc{G},\m{p}^*)$, where $\m{p}^* = [\m{p}^{*\top}_1,\ldots,\m{p}^{*\top}_n]^\top \in \mb{R}^{dn}$, is defined as follows:
\begin{Definition}[Desired formation] \label{def:target_formation}
The desired formation satisfies
\begin{enumerate}
    \item [(i)] Leaders' positions: $\m{p}_i^* = \m{p}_i,~\forall i=1,\ldots, l$, and
    \item[(ii)] Bearing constraints: $\m{g}_{ij}^* = \frac{\m{p}_j^*-\m{p}_i^*}{\|\m{p}_j^*-\m{p}_i^*\|},\; \forall (i,j)\in \mc{E}$.
\end{enumerate}
\end{Definition}

The following assumption will be used throughout the paper.
\begin{Assumption} \label{assumption} The formation of $l\ge 2$ leaders and $n-l$ followers satisfies
\begin{enumerate}
\item[(i)] The axes of the local reference frames of $n$ agents are aligned. 
\item[(ii)] The follower agents are modeled by \eqref{eq:agent_model} and there is no collision between agents.
\item[(ii)] The disturbance vector $\m{d}(t)=[\m{d}_1(t)^\top,\ldots,\m{d}_n(t)^\top]^\top$ is bounded and uniformly continuous. The direction and the upper bound of the disturbance (denoted as $\beta = \sup_{t\ge 0}\|\m{d}(t)\|_{\infty}>0$) are unknown to the agents. 
\item[(iii)] The target formation $(\mc{G},\m{p}^*)$ is infinitesimally bearing rigid in $\mb{R}^{d}$ and $l\ge 2$.
\end{enumerate}
\end{Assumption}

By stacking the set of desired bearing vectors as $\m{g}^* = [\ldots,(\m{g}_{ij}^*)^{\top},\ldots]^\top$, we have 
\begin{align} \label{eq:BearingLaplacian}
    \m{L}_b(\m{p}^*) \m{p}^*&= \m{L}_b(\m{g}^*) \m{p}^* \nonumber\\
    &=\begin{bmatrix}
        \m{L}_{ll}(\m{g}^*) & \m{L}_{lf}(\m{g}^*)\\
        \m{L}_{fl}(\m{g}^*) & \m{L}_{ff}(\m{g}^*)
    \end{bmatrix}\begin{bmatrix}
        \m{p}^{\rm L} \\ \m{p}^{\rm F*}
    \end{bmatrix} = \m{0}_{dn} \nonumber \\
    \Longrightarrow \m{L}_{fl}(\m{g}^*)\m{p}^{\rm L} &+ \m{L}_{ff}(\m{g}^*)\m{p}^{\rm F*} = \m{0}_{d(n-l)}.
\end{align}
Under the assumption that $(\mc{G},\m{p}^*)$ is infinitesimally bearing rigid in $\mb{R}^{d}$ and $l\geq 2$, it has been shown in \cite{Zhao2016aut} that $\m{L}_{ff}(\m{p}^*)$ is {symmetric positive definite and thus invertible. As a result,} the desired formation is uniquely determined from the the leaders' positions and the bearing vectors by $\m{p}^{\rm F*} = - (\m{L}_{ff}(\m{g}^*))^{-1}\m{L}_{fl}(\m{g}^*)\m{p}^{\rm L}$.

To achieve a target formation, the agents need to sense some geometric variables relating to the formation. Two types of relative sensing variables, namely, the displacements $\m{z}_{ij} = \m{p}_j - \m{p}_i,~\forall j \in \mc{N}_i$, and the bearing vectors $\m{g}_{ij} = \frac{\m{p}_j - \m{p}_i}{\|\m{p}_j - \m{p}_i\|}$, $\forall j \in \mc{N}_i$, will be considered in this paper. We can now formulate two problems which will be studied in the next sections.

\begin{Problem} \label{problem:1}
Let Assumption~\ref{assumption} hold and the agents can sense the relative displacements. Design control laws for agents such that $\m{p}(t) \to \m{p}^*$ as $t\to +\infty$.
\end{Problem}

\begin{Problem} \label{problem:2}
Let Assumption~\ref{assumption} hold and the agents can sense the bearing vectors.  Design control laws for agents such that $\m{p}(t) \to \m{p}^*$ as $t \to +\infty$.
\end{Problem}

\section{Displacement-based formation control}
\label{sec:3}
In this section, the bearing-based formation control under disturbance is considered under the assumption that the agents can measure the displacement vectors with regard to their neighbors. First, an adaptive variable structure control law which can provide asymptotic convergence of the target formation is proposed. Then, the proposed control law is modified to deal with different assumptions of the disturbances as well as the control objectives.
\subsection{Proposed control law}
Consider the Problem \ref{problem:1}, the control law is proposed as follows
\begin{subequations}
\label{eq:bearing_based_control_law}
\begin{align} 
    \m{u}_i &= - \sum_{j\in \mc{N}_i} \gamma_{ij} \m{P}_{\m{g}^*_{ij}} \text{sgn}(\m{q}_{ij}),  \, i=l+1,\ldots, n,\label{eq:bearing_based_control_law_a}\\
        \m{q}_{ij} &= {\m{P}_{\m{g}^*_{ij}}(\m{p}_i-\m{p}_j)}, \label{eq:bearing_based_control_law_b}\\
    \dot{\gamma}_{ij} &=\kappa \left|\left|\m{q}_{ij} \right|\right|_1,~\forall (i,j)\in \mc{E},\label{eq:bearing_based_control_law_c}
\end{align}
\end{subequations}
where, corresponding to each edge $e_k=(i,j)$, the matrix $\m{P}_{\m{g}^*_{ij}}=\m{I}_d-\m{g}^*_{ij}(\m{g}^*_{ij})^\top$ can be computed from the desired bearing vector $\m{g}^*_{ij} \equiv \m{g}_k^*$, the scalar $\gamma_{ij}$ are adaptive gains, which satisfy $\gamma_{ij}(0)>0$, and $\kappa>0$ is a positive constant. As the leaders are stationary, $\m{u}_i=\m{0}_d$ for $i=1,\ldots,l$. In the following analysis, let $\m{u}^{\rm L} = [\m{u}_{1}^\top,\ldots,\m{u}_l^\top]^\top = \m{0}_{dl}$, $\m{u}^{\rm F}=[\m{u}_{l+1}^\top,\ldots,\m{u}_n^\top]^\top$, $\m{u} = [\m{u}_{1}^\top,\ldots,\m{u}_n^\top]^\top = [(\m{u}^{\rm L})^\top,(\m{u}^{\rm F})^\top]^\top$, $\boldsymbol{\gamma}=[\ldots,\gamma_{ij},\ldots]^\top=[\gamma_1,\ldots,\gamma_m]^\top$, $\boldsymbol{\Gamma} = \text{diag}(\boldsymbol{\gamma})$, and $\bar{\boldsymbol{\Gamma}} = \boldsymbol{\Gamma} \otimes \m{I}_d$.

The system under the proposed control law \eqref{eq:bearing_based_control_law} can be expressed in the following form:
\begin{subequations}
\label{eq:bearing_based_system}
\begin{align}
    \dot{\m{p}} &=- \bar{\m{Z}} \left((\tilde{\m{R}}_b(\m{p}^*))^\top \bar{\boldsymbol{\Gamma}}\text{sgn} \left(\text{blkdiag}(\m{P}_{\m{g}^*_k})\bar{\m{H}}\m{p}\right)-\m{d} \right), \\
    \dot{\boldsymbol{\gamma}} &= \kappa \left[\|\m{q}_1\|_1,\ldots,\|\m{q}_m\|_1\right]^\top,
\end{align}
\end{subequations}
where $\m{Z} = \begin{bmatrix}
    \m{0}_{l\times l} & \m{0}_{l \times f} \\ \m{0}_{f \times l} & \m{I}_{f}
\end{bmatrix}$ and $\bar{\m{Z}} = \m{Z} \otimes \m{I}_d$. For brevity, the short-hands $\tilde{\m{R}}_b(\m{p}^*) = \tilde{\m{R}}_b(\m{g}^*)= \tilde{\m{R}}_b^*$, $\m{L}_b(\m{g}^*)=\m{L}_b^*$, and $\m{L}_{ff}(\m{g}^*)=\m{L}_{ff}^*$ will be used in the subsequent analysis.

\subsection{Stability analysis}
In this subsection, the stability of the system \eqref{eq:bearing_based_system} is considered. Since the right-hand-side of Eqn.~\eqref{eq:bearing_based_system}(a) is discontinuous, the solution of \eqref{eq:bearing_based_system}(a) is understood in Fillipov sense \cite{Shevitz1994}. It  will be proved that $\m{p}(t)$ converges to $\m{p}^*$ as $t\to +\infty$ under the proposed control law~\eqref{eq:bearing_based_control_law}.

\begin{Lemma} \label{lem:3.2}
Consider the Problem~\ref{problem:1}. If $\gamma_{ij}(0)>\gamma_0:=\beta \sqrt{\frac{dn}{\lambda_{\min}(\m{L}_{ff}^*)}}$, $\forall (i,j) \in \mc{E}$. Under the control law~\eqref{eq:bearing_based_control_law}, $\m{p}(t) \to \m{p}^*$ in finite time.
\end{Lemma}

\begin{IEEEproof}
Let $\boldsymbol{\delta} = \m{p} - \m{p}^{*} = [\m{0}_{dl}^\top,(\m{p}^{\rm F}-\m{p}^{\rm F*})^\top]^\top$, and consider the Lyapunov function $V = \frac{1}{2}\|\boldsymbol{\delta}\|^2$, which is positive definite, radially unbounded, and bounded by two class $\mc{K}_{\infty}$ functions $\frac{h_1}{2} \|\boldsymbol{\delta}\|^2$ and $\frac{h_2}{2} \|\boldsymbol{\delta}\|^2$, for any $0<h_1 < 1 < h_2$. Then, $\dot{V} \in^{a.e} \dot{\tilde{V}} = \bigcup_{\boldsymbol{\xi} \in \partial V} \boldsymbol{\xi}^\top \text{K}[\dot{\boldsymbol{\delta}}]$, where $\partial V = \{\boldsymbol{\delta}\}$ and $\in^{\rm a.e.}$ stands for almost everywhere \cite{Shevitz1994}. It follows that
\begin{align}
    \dot{{V}} &= \boldsymbol{\delta}^\top \bar{\m{Z}} \left(-(\tilde{\m{R}}_b^*)^\top\bar{\boldsymbol{\Gamma}}\text{K}[\text{sgn}] \big(\text{blkdiag}(\m{P}_{\m{g}^*_k})\m{z}\big)+\m{d}\right) \nonumber\\
    &= -\boldsymbol{\delta}^\top(\tilde{\m{R}}_b^*)^\top\bar{\boldsymbol{\Gamma}}\text{K}[\text{sgn}] \big(\text{blkdiag}(\m{P}_{\m{g}^*_k})\m{z}\big) + \boldsymbol{\delta}^\top\m{d} \nonumber \\
    &= -\m{z}^\top\text{blkdiag}\left(\m{P}_{\m{g}^*_k}\right)\bar{\boldsymbol{\Gamma}}\text{K}[\text{sgn}] \big(\text{blkdiag}(\m{P}_{\m{g}^*_k})\m{z}\big) + \boldsymbol{\delta}^\top\m{d} \nonumber\\
    &= -\sum_{k=1}^m \gamma_k(t) \|\m{P}_{\m{g}_k^*} \m{z}_k \|_1 + \boldsymbol{\delta}^\top\m{d} \nonumber \\
    &\leq -\underbrace{\min_{k}\gamma_{k}(0)}_{:={\chi}} \sum_{k=1}^m \|\m{P}_{\m{g}_k^*} \m{z}_k \|_1 + \|\boldsymbol{\delta}\|_1 \|\m{d}\|_{\infty}, \label{eq:4}
\end{align}
Note that in the third equality, we have used the fact that $(\m{p}^*)^\top (\tilde{\m{R}}_b^{*})^\top=\m{0}_{dn}^\top$, and the inequality \eqref{eq:4} follows from the fact that $\gamma_k(t)\ge \gamma_k(0) \ge \min_k \gamma_k(0) >0$. Based on the norm inequality for a vector $\m{x} \in \mb{R}^{dn}$, $\|\m{x}\| \leq \|\m{x}\|_1 \leq \sqrt{dn} \|\m{x}\|,$ we can further write
\begin{align}
    \dot{V} &\leq - {\chi}  \sum_{k=1}^m \|\m{P}_{\m{g}_k^*} \m{z}_k \|_1 + \|\boldsymbol{\delta}\|_1 \|\m{d}\|_{\infty} \nonumber\\
    &\leq -{\chi} \| \tilde{\m{R}}_{b}^* \m{p}\|_1 + \|\boldsymbol{\delta}\|_1 \|\m{d}\|_{\infty} \nonumber\\
    &\leq -{\chi} \| \tilde{\m{R}}_{b}^* \boldsymbol{\delta}\| + \|\boldsymbol{\delta}\|_1 \|\m{d}\|_{\infty} \nonumber\\
    &\leq - {\chi} \left(\boldsymbol{\delta}^\top \boldsymbol{\m{L}}_b^*\boldsymbol{\delta}\right)^{1/2} + \sqrt{dn}\|\boldsymbol{\delta}\| \|\m{d}\|_{\infty} \nonumber \\
    &= - {\chi} \left((\boldsymbol{\delta}^{\rm F})^\top \boldsymbol{\m{L}}_{ff}^*\boldsymbol{\delta}^{\rm F}\right)^{1/2} + \sqrt{dn} \beta \|\boldsymbol{\delta}\|. \label{eq:5}
\end{align}
Substituting the inequality $(\boldsymbol{\delta}^{\rm F})^\top \boldsymbol{\m{L}}_{ff}^* \boldsymbol{\delta}^{\rm F} \geq \lambda_{\min} (\boldsymbol{\m{L}}_{ff}^*) \|\boldsymbol{\delta}^{\rm F}\|^2 = \lambda_{\min} (\boldsymbol{\m{L}}_{ff}^*) \|\boldsymbol{\delta}\|^2$ into equation \eqref{eq:5}, we get
\begin{align}
    \dot{V} \leq - \underbrace{({\chi} \sqrt{\lambda_{\min}(\boldsymbol{\m{L}}_{ff}^*)} - \sqrt{dn} \beta)}_{:=\frac{1}{\sqrt{2}}\varepsilon} \|\boldsymbol{\delta}\| \leq - \varepsilon \sqrt{V}.\label{eq:ineq_V0}
\end{align}
We prove finite-time convergence of the desired formation by contradiction. If there does not exist a finite time $T>0$ such that $V(T)= 0$, and $V(t) = 0~\forall t \ge T$, then it follows from \eqref{eq:ineq_V0} that
\begin{align}
\frac{1}{2} \int_{V(0)}^{V(t)} \frac{dV}{\sqrt{V}} \leq - \frac{\varepsilon}{2} \int_0^t d\tau,
\end{align} 
or i.e., 
\begin{align} \label{eq:ineq_V}
0 \leq \sqrt{V(t)} \leq \sqrt{V(0)} - \frac{\varepsilon}{2}(t-0).
\end{align}
When $t>\frac{2\sqrt{V(0)}}{\varepsilon}$, the right hand side of the inequality \eqref{eq:ineq_V} becomes negative, which causes a contradiction. This contradiction implies that $\exists T>0: V(t)=0$ and for $t \geq T$. Thus, we conclude that $\m{p}(t) \to \m{p}^*$ in finite time. An upper bound for $T$ is thus $\frac{2\sqrt{V(0)}}{\varepsilon}$.
\end{IEEEproof}

Lemma \ref{lem:3.2} suggests that if initially, the gains $\gamma_{ij}$ have been chosen sufficiently large {(to dominate $\gamma_0$)}, the desired formation is achieved in finite time. However, some quantities such as the smallest eigenvalue of the grounded bearing Laplacian $\lambda_{\min}(\m{L}_{ff}^*)$ and the number of agents $n$ are usually unavailable. The proposed adaptive mechanism \eqref{eq:bearing_based_control_law_c2} makes the agents achieve the desired formation without requiring any a-priori information on the number of agents $n$, the desired formation's structure $\lambda_{\min}(\m{L}_{ff}^*)$ and the upper bound $\beta$ of the disturbance.

\begin{Theorem} \label{thm:1}
Consider the Problem \ref{problem:1}. Under the control law  \eqref{eq:bearing_based_control_law}, the following statements hold:
\begin{enumerate}[label = (\roman*)]
    \item $\m{p}(t) \to \m{p}^*$, as $t \to + \infty$, 
    \item There exists a constant vector $\boldsymbol{\gamma}^*=[\ldots,\gamma_{ij}^*,\ldots]^\top = [\gamma_1^*,\ldots,\gamma_m^*]$, such that $\boldsymbol{\gamma}(t) \to \boldsymbol{\gamma}^*$, as $t \to + \infty$,
    \item Additionally, if $\gamma_k^*> \gamma_0:=\beta \sqrt{\frac{dn}{\lambda_{\min}(\m{L}_{ff}^*)}},~\forall k=1,\ldots,m$, and there exists a finite time $T$ such that $|\gamma_k - \gamma_k^*|< \min_k|\gamma_k - \gamma_0|,~\forall i=1,\ldots,m,$ then $\m{p}(t) \to \m{p}^*$ in finite time.
\end{enumerate}
\end{Theorem}

\begin{IEEEproof}
(i) Consider the Lyapunov function $V = \frac{1}{2} \|\boldsymbol{\delta}\|^2 + \frac{1}{2\kappa} \|\boldsymbol{\gamma} - \bar{\gamma} \m{1}_m\|^2$, for some $\bar{\gamma} > \gamma_0$. $V$ is positive definite with regard to $\m{x}=[\boldsymbol{\delta}^\top,(\boldsymbol{\gamma}- \bar{\gamma} \m{1}_m)^\top]^\top$, radially unbounded, and bounded by two class $\mc{K}_{\infty}$ functions $h_1(\|\m{x}\|)=\min\{0.5,0.5\kappa^{-1}\}\|\m{x}\|^2$ and $h_2(\|\m{x}\|)=\max\{0.5,0.5\kappa^{-1}\}\|\m{x}\|^2$. Similar to the proof of Lemma~\ref{lem:3.2}, we have
\begin{align}
    \dot{V} &= -\sum_{k=1}^m \gamma_k(t) \|\m{P}_{\m{g}_k^*} \m{z}_k \|_1 + \boldsymbol{\delta}^\top\m{d} + \sum_{k=1}^m (\gamma_k - \bar{\gamma}) \|\m{P}_{\m{g}_k^*} \m{z}_k \|_1 \nonumber\\
    &\leq -\sum_{k=1}^m\bar{\gamma} \|\m{P}_{\m{g}_k^*} \m{z}_k \|_1 + \boldsymbol{\delta}^\top\m{d} \nonumber\\
    &\leq - (\bar{\gamma}\sqrt{\lambda_{\min}(\boldsymbol{\m{L}}_{ff}^*)} - \sqrt{dn}\beta) \| \boldsymbol{\delta}\| \leq 0,
\end{align}
which implies that $\boldsymbol{\delta}$, $\boldsymbol{\gamma}$ are bounded, and $\exists \lim_{t \to + \infty} V(t) \geq 0$. Since $\dot{V}$ is uniformly continuous, it follows from Barbalat's lemma that $\dot{V} \to {0}$, or $\boldsymbol{\delta} \to \m{0}_{dn}$, as $t \to \infty$. 

(ii) Since $-\gamma_k, ~k=1,\ldots, m,$ is bounded and non-increasing, it has a finite limit. Thus, there exists $\boldsymbol{\gamma}^*$ such that $\boldsymbol{\gamma}(t) \to \boldsymbol{\gamma}^*$, as $t \to + \infty$. 

(iii) If there exists a finite time $T$ such that $|\gamma_k - \gamma_k^*|< \min_k|\gamma_k - \gamma_0|,~\forall i=1,\ldots,m,$ then for all $t\geq T$, the inequality \eqref{eq:ineq_V0} holds. Therefore, the proof of this statement follows directly from the proof of Lemma \ref{lem:3.2}.
\end{IEEEproof}

\begin{Remark} \label{rem:0} Since the control law \eqref{eq:bearing_based_control_law} uses only signum functions, chattering will be unavoidable. To reduce the magnitude of chattering, a proportional control term $- k_p \sum_{j\in \mc{N}_i} \gamma_{ij} \m{P}_{\m{g}^*_{ij}} \m{q}_{ij}$ into (\ref{eq:bearing_based_control_law}a) can be included as follows:
\begin{align*}
\m{u}_i = - k_p \sum_{j\in \mc{N}_i} \gamma_{ij} \m{P}_{\m{g}^*_{ij}} \m{q}_{ij} - \sum_{j\in \mc{N}_i} \gamma_{ij} \m{P}_{\m{g}^*_{ij}} {\rm sgn}(\m{q}_{ij}),  \,
\end{align*}
for $i=l+1,\ldots, n$. If there is no disturbance, the proportion control term is sufficient for achieving the target formation. When disturbances exist, the proportion term contributes to the formation acquisition and disturbance rejection objectives, at a slower rate in comparison with the signum term.
\end{Remark}

\begin{Remark} \label{rem:2} An issue with the control law \eqref{eq:bearing_based_control_law_a}--\eqref{eq:bearing_based_control_law_c} is that the control gains $\gamma_{ij}$ is non-decreasing at any time $t\geq 0$. Thus, if the disturbance has a high magnitude for a time interval, and then decreases in time, much control effort will be wasted. To address this issue, we may relax the objective from perfectly achieving a target formation into achieving a good approximation of the target formation. More specifically, we may control the formation under disturbances to reach a small neighborhood of the desired formation in finite-time while the control magnitude estimates the unknown upper bound of the disturbance \cite{Oliveira2016adaptive}. A corresponding modified formation control law is then {proposed} as follows:
\begin{subequations}
\label{eq:bearing_based_control_law2}
\begin{align}
    \m{u}_i &= - k_p\sum_{j\in \mc{N}_i} \m{P}_{\m{g}^*_{ij}} \m{z}_{ij} - \sum_{j\in \mc{N}_i} \gamma_{ij} \m{P}_{\m{g}^*_{ij}} {\rm sgn}(\m{q}_{ij}), \label{eq:bearing_based_control_law_a2}\\
        \m{q}_{ij} &= {\m{P}_{\m{g}^*_{ij}}(\m{p}_i-\m{p}_j)} \equiv \m{q}_k, \label{eq:bearing_based_control_law_b2}\\
    \dot{\gamma}_{ij} &= \kappa(\|\m{q}_{ij} \|_1 - \alpha \gamma_{ij}) \equiv \dot{\gamma}_k,\\
     \gamma_{ij}&(0)>0,\,\forall e_k = (i,j) \in \mc{E},
 \label{eq:bearing_based_control_law_c2}
\end{align}
\end{subequations}
where $i=1,\ldots, n$, and $\alpha, \kappa>0$ are positive constants. For each $(i,j)\in \mc{E}$,
\begin{align*}
{\gamma}_{ij}(t) = \underbrace{e^{-\kappa\alpha t} \gamma_{ij}(0)}_{\geq 0} + \underbrace{\kappa\int_0^{t} e^{-\kappa\alpha (t-\tau)} \|\m{q}_{ij} \|_1 d\tau}_{\geq 0} \geq 0,\, \forall t\geq 0.
\end{align*}

Similar to the proof of Theorem \ref{thm:1}, consider the Lyapunov function $V = \frac{1}{2} \|\boldsymbol{\delta}\|^2 + \frac{1}{2\kappa} \|\boldsymbol{\gamma} - \bar{\gamma} \m{1}_m\|^2$, where $\bar{\gamma} > \gamma_0$. We have,
\begin{align}
    \dot{V} &= -\sum_{k=1}^m \left(k_p\|\m{P}_{\m{g}_k^*} \m{z}_k \|^2 + \gamma_k \|\m{q}_k \|_1 \right) + \boldsymbol{\delta}^\top \m{d} \nonumber\\
    &\qquad\qquad\qquad + \sum_{k=1}^m (\gamma_k - \bar{\gamma}) (\|\m{q}_k \|_1 - \alpha \gamma_{k}) \nonumber\\
    &\leq -\sum_{k=1}^m \left(k_p \|\m{q}_k \|^2 + \alpha(\gamma_k^2 - \bar{\gamma}\gamma_k) \right)  - \bar{\gamma}\sum_{k=1}^m \|\m{q}_k\|_1 + \boldsymbol{\delta}^\top \m{d} \nonumber \\ 
    &\leq -\sum_{k=1}^m \left(k_p \|\m{q}_k \|^2 + \frac{1}{2}\alpha(2\gamma_k^2 - 2\bar{\gamma}\gamma_k + \bar{\gamma}^2) - \frac{1}{2}\alpha \bar{\gamma}^2 \right) \nonumber \\
    &\qquad\qquad - \left(\bar{\gamma}\sqrt{\lambda_{\min}(\boldsymbol{\m{L}}_{ff}^*)} - \sqrt{dn}\beta\right) \| \boldsymbol{\delta}\| \nonumber\\
    &\leq -\sum_{k=1}^m \left(k_p \|\m{q}_k \|^2 + \frac{1}{2}\alpha(\gamma_k - \bar{\gamma})^2 \right) + \frac{m}{2}\alpha \bar{\gamma}^2 \nonumber\\
    &\leq - k_p \boldsymbol{\delta}^\top \m{L}_b^* \boldsymbol{\delta} - \frac{\alpha}{2} \|\boldsymbol{\gamma}- \bar{\gamma} \m{1}_m\|^2 + \frac{m}{2}\alpha \bar{\gamma}^2 \nonumber\\
    &\leq -k_p \lambda_{\min}(\boldsymbol{\m{L}}_{ff}^*) \|\boldsymbol{\delta}\|^2 - \frac{\alpha}{2} \|\boldsymbol{\gamma}- \bar{\gamma} \m{1}_m\|^2 + \frac{m}{2}\alpha \bar{\gamma}^2. \label{eq:rem1}
\end{align}
Let $\varrho = \min\{2 k_p \lambda_{\min}(\boldsymbol{\m{L}}_{ff}^*),\kappa\alpha \}$, we have,
\begin{align}
\dot{V} &\leq -\varrho V + \frac{m}{2}\alpha \bar{\gamma}^2 \nonumber \\
&= -\varrho (1-\theta) V -\varrho\theta V + \frac{m}{2}\alpha \bar{\gamma}^2,
\end{align}
for some $\theta\in (0,1)$. Thus, when $V(t)\ge \Delta:=\frac{m}{2\varrho\theta}\alpha\bar{\gamma}^2$, we have $\dot{V}\leq - \varrho (1-\theta)V$, or $V\leq \max\{V(0),\Delta\}$. Thus, $\m{x}=[\boldsymbol{\delta}^\top,(\boldsymbol{\gamma} - \bar{\gamma} \m{1}_m)^\top]^\top$ is globally ultimately bounded. Defining the ball $\mc{B}_{\Delta} = \{ \m{x}=[\boldsymbol{\delta}^\top,(\boldsymbol{\gamma} - \bar{\gamma} \m{1}_m)^\top]^\top\in \mb{R}^{dn+m}|~\|\m{x}\|\leq h_2^{-1}(h_1(\Delta))\}$, then $\m{x}$ enters the ball $\mc{B}_{\Delta}$ after a finite time. It follows that  $\|\boldsymbol{\delta}\|=\|\m{p}(t)-\m{p}^*\|\leq h_2^{-1}(h_1(\Delta))$ after a finite time.

It is worth noting that by relaxing the control objective, we also further reduce the chattering behaviors of the formation in both magnitude and switching frequency. Most control efforts are provided to maintain the formation error inside a closed ball, of which the radius is jointly determined by the desired formation (number of bearing constraints $m$ and the minimum eigenvalue $\lambda_{\min}(\boldsymbol{\m{L}}_{ff}^*$) and the control parameters (proportional control gain $k_p$, adaptation rate $\kappa$, and the decay rate $\alpha$). Other methods for avoiding chattering may be softening the sign function by the tanh($\cdot$) function \cite{Garanayak2023bearing}, or considering a deadzone once error is small enough. Nevertheless, all above mentioned methods need to sacrifice the control performance for eradication of chattering.
\end{Remark}

In the next remark, we further consider a larger class of the disturbance acting on the formation. Let the upper bound of the disturbance be a polynomials of the formation's error. The main idea is to design adaptive law for each coefficient term \cite{Roy2020adaptive}.

\begin{Remark} \label{rem:3} Suppose that the upper bound of the unknown disturbance acting on the formation satisfies 
\begin{align}
\|\m{d}(t)\|_{\infty} &\leq \beta_0 + \beta_1 \| \boldsymbol{\delta} \|_1+ \ldots + \beta_N \|\boldsymbol{\delta}\|_1^N  = \sum_{r=0}^N \beta_r \|\boldsymbol{\delta}\|_1^r,\nonumber
\end{align}
$\forall t\geq 0, \mb{N}_{+} \ni N$, where $\beta_1,\ldots,\beta_N$ are unknown positive constants.

The following adaptive formation control law is proposed:
\begin{subequations}
\label{eq:bearing_based_control_law1}
\begin{align} 
    \m{u}_i &= - \sum_{j\in \mc{N}_i} \gamma_{ij}(t) \m{P}_{\m{g}^*_{ij}} {\rm sgn}(\m{q}_{ij}), \, i=1,\ldots, n, \label{eq:bearing_based_control_law_a1}\\
        \m{q}_{ij} &= {\m{P}_{\m{g}^*_{ij}}(\m{p}_i-\m{p}_j)} \equiv \m{q}_{k}, \label{eq:bearing_based_control_law_b1}\\
        {\gamma}_{ij}(t) &= \hat{\beta}_0^{ij}(t) + \hat{\beta}_1^{ij}(t) \|\m{q}_{ij}\|_1 + \ldots + \hat{\beta}_N^{ij}(t) \|\m{q}_{ij}\|_1^N \equiv \gamma_k(t) \\
        \hat{\beta}_r^{ij}(t) &\equiv \hat{\beta}_1^{k}(t),\,\forall e_k=(i,j) \in \mc{E},~k = 1, \ldots, m.\\
    \dot{\hat{\beta}}_0^{ij}(t) &= \left|\left|\m{q}_{ij} \right|\right|_1,\,{\hat{\beta}}_0^{ij}(0)>0, \label{eq:bearing_based_control_law_c1} \\
    \dot{\hat{\beta}}_1^{ij}(t) &= \left|\left|\m{q}_{ij} \right|\right|_1^2,\,{\hat{\beta}}_1^{ij}(0)>0,  \label{eq:bearing_based_control_law_d1}\\
    &\vdots \nonumber\\
     \dot{\hat{\beta}}_N^{ij}(t) &= \left|\left|\m{q}_{ij} \right|\right|_1^{N+1},\,{\hat{\beta}}_N^{ij}(0)>0. \label{eq:bearing_based_control_law_e1}
\end{align}
\end{subequations}
For stability analysis, let $\boldsymbol{\hat{\beta}}_r = [\ldots,\hat{\beta}_r^{ij},\ldots]^\top = [\hat{\beta}_r^{1},\ldots,\hat{\beta}_r^{m}]^\top\in \mb{R}^m$, $\forall r=0,1,\ldots,N$, and consider the Lyapunov candidate function \[V = \frac{1}{2} \|\boldsymbol{\delta}\|^2 + \frac{1}{2} \sum_{r=0}^N\|\boldsymbol{\hat{\beta}}_r - \bar{\beta}_r \m{1}_m\|^2,\] where $\bar{\beta_r}>\beta_r \left(\sqrt{\frac{\lambda_{\min}(\boldsymbol{\m{L}}_{ff}^*)}{dn}}\right)^{r+1}, \forall r = 0, \ldots, N$. Then,
\begin{align}
\dot{V} &= -\sum_{k=1}^m \gamma_{k} \|\m{q}_{k} \|_1 + \boldsymbol{\delta}^\top \m{d} + \sum_{k=1}^m\sum_{r=0}^N (\hat{\beta}_{r}^k - \bar{\beta}_{r}) \|\m{q}_k \|_1^{r+1} \nonumber\\
    & \leq  -\sum_{k=1}^m \sum_{r=0}^N \hat{\beta}_{r}^k \|\m{q}_k\|_1^{r+1} + \|\boldsymbol{\delta}\|_1 \|\m{d}\|_{\infty}  \\
    &\qquad \quad + \sum_{k=1}^m\sum_{r=0}^N (\hat{\beta}_{r}^k - \bar{\beta}_{r}) \|\m{q}_k \|_1^{r+1}. \nonumber
\end{align}
It follows that
\begin{align}
 \dot{V}   &\leq - \sum_{r=0}^N \bar{\beta}_{r} \sum_{k=1}^m \|\m{q}_k \|_1^{r+1} + \sum_{r=0}^N \beta_r \|\boldsymbol{\delta}\|_1^{r+1}\nonumber\\
    &\leq - \sum_{r=0}^N \bar{\beta}_{r} \left(\sum_{k=1}^m \|\m{q}_k \|_1\right)^{r+1} + \sum_{r=0}^N \beta_r \|\boldsymbol{\delta}\|_1^{r+1}\nonumber\\
    &\leq - \sum_{r=0}^N \bar{\beta}_{r} \left(\sqrt{\lambda_{\min}(\boldsymbol{\m{L}}_{ff}^*)} \|\boldsymbol{\delta}\| \right)^{r+1}  + \sum_{r=0}^N \beta_r \left({dn}\right)^{\frac{r+1}{2}} \|\boldsymbol{\delta}\|^{r+1} \nonumber\\
    & \leq -\sum_{r=0}^N \underbrace{\left(\bar{\beta}_{r}\left(\sqrt{\lambda_{\min}(\boldsymbol{\m{L}}_{ff}^*)} \right)^{r+1} - \beta_r\left({dn}\right)^{\frac{r+1}{2}}\right)}_{>0} \|\boldsymbol{\delta}\| ^{r+1} \nonumber\\
    &\leq 0.
\end{align}
It follows that $\boldsymbol{\delta}$ and $\hat{\boldsymbol{\beta}}_r,~\forall r=0,1,\ldots,N,$ are uniformly bounded. Similar to the proof of Theorem \ref{thm:1}, we can show that $\|\boldsymbol{\delta}\| \to \m{0}_{dn}$, or $\m{p}(t) \to \m{p}^*$, as $t\to \infty$, and $\lim_{t\to\infty}\hat{\boldsymbol{\beta}}_r,~\forall r=0,1,\ldots,N,$ exists. Further, if  $\hat{\boldsymbol{\beta}}_r > \bar{\beta}_r\m{1}_m,~\forall r=0,1,\ldots,N$, where ``$>$'' is understood to be element-wise, then $\m{p}(t)\to \m{p}^*$ in finite time.
\end{Remark}

\section{Bearing-only based formation control}
\label{sec:4}
In this section, we further assume that the agents can measure only the relative bearing vectors with regard to their neighbors. We propose a corresponding adaptive variable-structure bearing-only formation control law and showed that the desired formation can be asymptotically achieved. Moreover, due to the adaptive gains, the effects of unknown time-varying disturbances acting on formation can be completely rejected even when the followers agents are not given any information of the disturbances' upper bound.
\subsection{Proposed control law}
Consider the system of single-integrator agents with disturbance \eqref{eq:agent_model}. The  bearing-only control law for each follower agent $i \in \{l+1,\ldots,n\}$ is proposed as follows
\begin{subequations}
\label{eq:Bearing_OnlyC}
\begin{align} 
    \m{u}_i &= - \gamma_{i} \text{sgn} \left({\m{r}_i} \right), ~{\m{r}_i} = -\sum_{j\in \mc{N}_i} (\m{g}_{ij} - \m{g}_{ij}^*), \label{eq:Bearing_OnlyC1}\\
    \dot{\gamma}_{i} &= \kappa_i \|{\m{r}_i}\|_1. \label{eq:Bearing_OnlyC2}
\end{align}
\end{subequations}
{Denoting $\m{r}=[\m{r}_1^\top, \ldots, \m{r}_n^\top]^\top = \bar{\m{H}}^\top (\m{g} - \m{g}^*)$,} we can express the $n$-agent system under the control law \eqref{eq:Bearing_OnlyC1}--\eqref{eq:Bearing_OnlyC2} in vector form as follows:
\begin{subequations}
\begin{align}
    \dot{\m{p}} &= \bar{\m{Z}} \left(-\bar{\boldsymbol{\Gamma}} \text{sgn}\left( \m{r} \right) + \m{d}\right), \label{eq:system_BOM}\\
    \dot{\boldsymbol{\gamma}} &= {\left[\ldots,\kappa_i\|{\m{r}_i} \|_1,\ldots\right]^\top}, \label{eq:system_BOM1}
\end{align}
\end{subequations}
where $\boldsymbol{\gamma} = [\gamma_1,\ldots,\gamma_{n}]^\top \in \mb{R}^n$, $\boldsymbol{\Gamma} = \text{diag}(\boldsymbol{\gamma})$ and $\bar{\boldsymbol{\Gamma}} = \boldsymbol{\Gamma} \otimes \m{I}_d$. It is clear that the control law of each agent uses only the bearing vectors with regard to its neighboring agents.

\subsection{Stability analysis}
This subsection studies the stability of the $n$-agent system \eqref{eq:system_BOM}--\eqref{eq:system_BOM1}. Particularly, we show that the desired formation $\m{p}^*$ defined as in Definition \ref{def:target_formation} will be asymptotically achieved as $t \to \infty$. Since the right-hand-side of Eq.~\eqref{eq:system_BOM} is discontinuous, we understand the solution of \eqref{eq:system_BOM} in Fillipov sense. 

We will firstly prove the following lemma.
\begin{Lemma}\cite[Lemma 2]{zhao2019bearing} \label{lem:2} Suppose that no agents coincide in $\m{p}$ or $\m{p}^*$. The following inequality holds
\begin{align}
\m{p}^\top {\m{r}} &\geq 0, \label{eq:19}\\
(\m{p}^*)^\top {\m{r}} &\leq 0, \label{eq:20}\\
(\m{p}-\m{p}^*)^\top {\m{r}} &\geq 0, \label{eq:21}
\end{align}
where the equality holds if and only if $\m{g}=\m{g}^*$.
\end{Lemma}

\begin{Lemma}\cite[Lemma 3]{zhao2019bearing} \label{lem:3} Suppose that no agents coincide in $\m{p}$ or $\m{p}^*$, then
\begin{align} \label{eq:lem5a}
\m{p}^\top {\m{r}} &\geq \frac{1}{2\max_{(i,j)\in \mc{E}} \|\m{z}_{ij}\|} \m{p}^\top \m{L}_b^*\m{p}.
\end{align}
Furthermore, if $\m{g}_k^\top\m{g}^*_k\geq 0,~\forall k =1,\ldots, m,$ then
\begin{align} \label{eq:lem5b}
\m{p}^\top {\m{r}} &\leq \frac{1}{\min_{(i,j)\in \mc{E}} \|\m{z}_{ij}\|} \m{p}^\top \m{L}_b^*\m{p}.
\end{align}
\end{Lemma} 

Next, we prove that the adaptive bearing-only control law \eqref{eq:Bearing_OnlyC} guarantees boundedness of the formation's error $\boldsymbol{\delta}=\m{p}-\m{p}^*$ in the following lemma.
\begin{Lemma} \label{lem:4}
Consider the Problem \ref{problem:2} {and suppose that there is no collision between agents for $t\geq 0$}. Under the control law \eqref{eq:Bearing_OnlyC}, the formation error $\boldsymbol{\delta} = \m{p} - \m{p}^*$ is uniformly bounded, {$\m{p}\to \m{p}^*$ as $t\to+\infty$ and there exists constant vector $\boldsymbol{\gamma}^*$ such that $\lim_{t\to +\infty}\boldsymbol{\gamma}(t) = \boldsymbol{\gamma}^*$.}
\end{Lemma}
\begin{IEEEproof}
Consider the Lyapunov function 
\begin{equation} \label{eq:Lyap_bom}
V = \m{p}^\top {\m{r}}  + \sum_{i=l+1}^n \frac{(\gamma_i -\gamma_0)^2}{2\kappa_i},
\end{equation}
 where $\gamma_0>\|\m{d}\|_{\infty}$. Then, $V=0$ if and only if $\m{p}^\top {\m{r}}=0$, or equivalently, $\m{g}=\m{g}^*$ and $\gamma_i =\gamma_0,\forall i=l+1,\ldots,n$. Since $(\mc{G},\m{p}^*)$ is infinitesimally rigid and $l\ge 2$, the equality $\m{g}=\m{g}^*$ implies that $\m{p}=\m{p}^*$. 
 We have
\begin{align}
\dot{V} &\in^{\text{a.e}} \dot{\tilde{V}} \nonumber\\
    &= -{\m{r}}^\top \bar{\m{Z}} \left(\bar{\boldsymbol{\Gamma}} {{\rm K}[\text{sgn}]}\left({\m{r}} \right) - \m{d}\right) + \sum_{i=1}^n (\gamma_i -\gamma_0) \|{\m{r}}_i\|_1 \nonumber\\
    &= -\sum_{i=l+1}^n \gamma_i \|{\m{r}_i}\|_1 - 
\sum_{i=l+1}^n {\m{r}}_i^\top \m{d}_i + \sum_{i=1}^n (\gamma_i -\gamma_0) \|{\m{r}}_i\|_1 \nonumber\\ 
    &\leq - \sum_{i=l+1}^n (\gamma_0 - \|\m{d}_i\|_{\infty}) \|{\m{r}_i}\|_1 \nonumber\\
    &\leq - (\gamma_0 - \|\m{d}\|_{\infty}) \left| \left| \sum_{i=l+1}^n{\m{r}}_i \right| \right|_1\nonumber\\
    &\leq - (\gamma_0 - \|\m{d}\|_{\infty}) \left|\left|\bar{\m{Z}} {\m{r}}\right|\right|_1\leq 0. \label{eq:pf1}
\end{align}
It follows that $V(t)\leq V(0),~\forall t\geq 0$,  $\m{z}^\top(\m{g}-\m{g}^*)$ and $(\boldsymbol{\gamma} - \gamma_0\m{1}_n)$ are always bounded. 

Further, from the inequalities
{
\begin{align}
    \|\m{z}_{ij}\| &= \|\m{p}_i-\m{p}_i^*+\m{p}_i^*-\m{p}_j^*+\m{p}_j^*-\m{p}_j\| \nonumber\\
    &\leq \|\bm{\delta}_i\|+\|\bm{\delta}_j\|+\|\m{p}_i^*-\m{p}_j^*\|  \nonumber\\
    &\leq \sum_{i=1}^n\|\bm{\delta}_i\| + \max_{i,j\in \mc{V},~i\neq j} \|\m{p}_i^*-\m{p}_j^*\|  \nonumber\\
    &\leq \sqrt{n} \|\bm{\delta}\| + \underbrace{\max_{i,j\in \mc{V},~i\neq j} \|\m{p}_i^*-\m{p}_j^*\|}_{:=M},~\forall (i,j)\in \mc{E}, \label{eq:26}
\end{align}
\begin{align}
    \lambda_{\min}(\m{L}_{ff}^*) \|\bm{\delta}\|^2\leq \m{p}^\top\m{L}_{b}^*\m{p} 
    \leq \lambda_{\max}(\m{L}_{ff}^*) \|\bm{\delta}\|^2,~\label{eq:27}
\end{align}}
and Eqn.~\eqref{eq:lem5a}, we have
\begin{align*}
V \geq \m{p}^\top{\m{r}} &\geq \frac{1}{2\max_{(i,j)\in \mc{E}} \|\m{z}_{ij}\|}\m{p}^\top \m{L}_b^*\m{p} \\
&\geq {\underbrace{\frac{\lambda_{\min}(\m{L}_{ff}^*)}{2}}_{:=\vartheta} }\frac{\|\boldsymbol{\delta}\|^2}{{\sqrt{n}}\|\boldsymbol{\delta}\| + {M}},
\end{align*}
which shows that ${\vartheta}\frac{\|\boldsymbol{\delta}\|^2}{{\sqrt{n}}\|\boldsymbol{\delta}\| + {M}}$ is bounded. Suppose that {$\alpha=V(0)\geq V(t)\geq 0$}, i.e., 
\begin{align*}
{\vartheta}\frac{\|\boldsymbol{\delta}\|^2}{{\sqrt{n}}\|\boldsymbol{\delta}\| + {M}} \leq \alpha.
\end{align*}
Let $x=\|\boldsymbol{\delta}\|\geq 0$ and $\zeta = \frac{\alpha}{\vartheta}>0$, the inequality
\[x^2-\zeta{\sqrt{n}} x-\zeta{M} \leq 0\]
has solution
\begin{align}
{0\leq }\|\boldsymbol{\delta}(t)\| \leq \frac{\zeta+ \sqrt{{n}\zeta^2 + 4{M}\zeta}}{2},~\forall t\geq 0. \label{eq:bounded_delta}
\end{align}
Thus, $\boldsymbol{\delta}(t)$ is uniformly bounded. As there is no collision between agents, $\dot{V}$ is uniformly continuous, and thus $\m{g}\to \m{g}^*$ as $t\to +\infty$ based on Barbalat's lemma. This implies that $\m{p}\to\m{p}^*$ as $t\to +\infty$. Moreover, since $(\gamma_i(t)-\gamma_0),~\forall i=l+1,\ldots,n,$ are bounded and nonincreasing, $\m{z}^\top(\m{g}-\m{g}^*) \to 0$, it follows that there exists $\boldsymbol{\gamma}^*$ such that $\lim_{t\to +\infty}\boldsymbol{\gamma}(t) = \boldsymbol{\gamma}^*$.
\end{IEEEproof}

The following lemma gives a sufficient condition for collision avoidance between neighboring agents.
\begin{Lemma} \label{lem:5} Consider the Problem \ref{problem:2}. Suppose that $0<\eta:=\frac{\zeta+ \sqrt{{n}\zeta^2 + 4{M}\zeta}}{2} < \frac{{N}}{\sqrt{n}}$, where ${N}:=\min_{i,j \in \mc{V},~{i\ne j}}\|\m{p}_i^*-\m{p}_j^*\|$, 
then $\min_{i,j\in \mc{V}}\|\m{p}_i-\m{p}_j\|\geq {N} - \sqrt{n} \eta,~\forall t\geq 0$.
\end{Lemma}
\begin{IEEEproof}
For each $i,j\in \mc{V},~i\ne j$, we can write $\m{p}_i-\m{p}_j = (\m{p}_i - \m{p}_i^*)- (\m{p}_j-\m{p}_j^*)+ (\m{p}_i^*-\m{p}_j^*)$. Thus,
\begin{align}
\|\m{p}_i-\m{p}_j\| &\geq \|\m{p}^*_i-\m{p}_j^*\| - \|\m{p}_j - \m{p}_j^*\| - \|\m{p}_i - \m{p}_i^*\| \nonumber\\
&\geq \|\m{p}^*_i-\m{p}_j^*\| - \sum_{i=1}^n\|\m{p}_i - \m{p}_i^*\| \nonumber\\
&\geq \|\m{p}^*_i-\m{p}_j^*\| - \sqrt{n}\|\boldsymbol{\delta}\|. \label{eq:29}
\end{align}
It follows from \eqref{eq:bounded_delta} that $\|\boldsymbol{\delta}(t)\|\leq \eta,~\forall t\geq 0$. Thus, we have
\[\min_{i,j\in\mc{V},{i\ne j}}\|\m{p}_i-\m{p}_j\| \geq {N} - \sqrt{n} \eta{>0},~\forall t\geq 0,\]
or in other words, no collision happens for all $t\geq 0$.
\end{IEEEproof}

In the following theorem, a sufficient condition for stabilizing the desired target formation will be given. 
\begin{Theorem} \label{thm:2}
Consider the Problem \ref{problem:2}. Under the adaptive bearing-only control law~\eqref{eq:Bearing_OnlyC}, there exists a positive constant $\alpha>0$ such that if the Lyapunov function in \eqref{eq:Lyap_bom} satisfies $V(0)<\alpha$, then $\min_{i,j\in \mc{V}{,i\neq j}}\|\m{p}_i-\m{p}_j\|\geq \theta>0,$ $\forall t\geq 0$,  $\lim_{t\to +\infty}\m{p}(t)= \m{p}^*$, and there exists $\boldsymbol{\gamma}^*$ such that $\lim_{t\to +\infty}\boldsymbol{\gamma}(t) = \boldsymbol{\gamma}^*$.
\end{Theorem}

\begin{IEEEproof}
From Eqn.~\eqref{eq:bounded_delta}, we obtain
\begin{align}
\|\boldsymbol{\delta}\| &\leq \frac{\alpha/\vartheta + \sqrt{{n}\alpha^2/\vartheta^2 + 4{M}(\alpha/\vartheta) \|\m{p}^*\|} }{2},~\forall t\geq 0. \label{eq:30}
\end{align} 
Thus, for $\alpha$ sufficiently small, the inequality $0<\eta=\frac{\zeta+ \sqrt{{n}\zeta^2 + 4{M}\zeta}}{2} <\frac{1}{\sqrt{n}}\min_{i,j \in \mc{V}{,i\neq j}}\|\m{p}_i^*-\m{p}_j^*\| {= \frac{N}{\sqrt{n}}}$ can always be satisfied. It follows from Lemma~\ref{lem:5} that no collision can happen, and $\min_{i,j\in\mc{V}{,i\neq j}}\|\m{p}_i-\m{p}_j\| \geq \theta:= {N} - \sqrt{n} \eta,~\forall t\geq 0.$ {As collision avoidance is ensured, $\m{g}_{ij},~\forall (i,j)\in \mc{E}$, are uniformly continuous, and so is $\dot{V}$. The remaining proof is similar to the proof of Lemma~\ref{lem:4} and will be omitted.}
\end{IEEEproof}

{
\begin{Remark} \label{rem:4} Similar to Remark~\ref{rem:2}, we may relax the objective from perfectly achieving a target formation into achieving a good approximation of the target formation with the following modified bearing-only formation control law
\begin{subequations}
\label{eq:bearing_only_control_law2}
\begin{align}
    \m{u}_i &= - k_p\m{r}_i  - \gamma_i(t){\rm sgn}\left(\m{r}_i\right),~ \label{eq:bearing_only_control_law_b2}\\
    \dot{\gamma}_{i} &= \kappa(\|\m{r}_{i} \|_1 - \alpha \gamma_{i}),\,
     \gamma_{i}(0)>0,\forall i=1,\ldots,n,
 \label{eq:bearing_only_control_law_c2}
\end{align}
\end{subequations}
where $\m{r}_{i} = \sum_{j\in \mc{N}_i}(\m{g}_{ij} - \m{g}^*_{ij})$ and $\alpha, \kappa>0$. Clearly, 
${\gamma}_{i}(t) \geq 0,\, \forall t\geq 0.$ Denoting $\m{r}=[\m{r}_1^\top,\ldots,\m{r}_n^\top]^\top$, similar to the proof of Lemma~\ref{lem:4} and the analysis in Remark~\ref{rem:2}, consider the Lyapunov function $V = \m{p}^\top\m{r} + \frac{1}{2\kappa} \sum_{i=l+1}^n(\gamma_i - \gamma_0)^2$, where $\gamma_0 > \|\m{d}\|_{\infty}$. We have,
\begin{align}
    \dot{V} &= -\sum_{i=l+1}^n\|\m{r}_i\|^2 - \sum_{i=l+1}^n \gamma_i\|\m{r}_i\|_1 + \sum_{i=l+1}^n \m{r}_i^\top\m{d}_i\nonumber\\
    &\qquad + \sum_{i=l+1}^n(\gamma_i-\gamma_0)(\|\m{r}_i\|_1 - \alpha\gamma_i) \nonumber\\
    &\leq -\sum_{i=l+1}^n\|\m{r}_i\|^2 - (\gamma_0 - \|\m{d}\|_{\infty})\sum_{i=l+1}^n\|\m{r}_i\|_1 \nonumber\\
    &\qquad -\frac{1}{2}\alpha\sum_{i=l+1}^n(\gamma_i - \gamma_0)^2 + \frac{n-l}{2}\alpha\gamma_0^2 \nonumber\\
    &\leq -k_p \|\bar{\m{Z}}\m{r}\|^2 - \alpha\sum_{i=l+1}^n\frac{(\gamma_i - \gamma_0)^2}{2} + \frac{n-l}{2}\alpha\gamma_0^2. \label{eq:rem4_ineq}
\end{align}
The Cauchy-Schwartz inequality $\|\m{a}\|\|\m{b}\|\geq |\m{a}^\top\m{b}|$ gives
\begin{align*}
   \|\bar{\m{Z}}\m{r}\|^2 &=
    \frac{\|\bm{\delta}^\top\|^2\|\bar{\m{Z}}\m{r}\|^2}{\|\bm{\delta}^\top\|^2} \geq \frac{\left(\bm{\delta}^\top\bar{\m{Z}}\m{r}\right)^2}{\|\bm{\delta}\|^2} = \frac{\left(\bm{\delta}^\top\m{r}\right)^2}{\|\bm{\delta}\|^2}.
\end{align*}
Next, using Lemma~\ref{lem:2}, the inequalities \eqref{eq:27}, and Lemma~\ref{lem:3}, there holds
\begin{align*}
\|\bar{\m{Z}}\m{r}\|^2  &\geq \frac{\left(\m{p}^\top\bar{\m{H}}^\top(\m{g}-\m{g}^*) - \m{p}^{*\top}\bar{\m{H}}^\top(\m{g}-\m{g}^*) \right)^2}{\|\bm{\delta}\|^2}\\
    &\geq \frac{\left(\m{p}^\top\bar{\m{H}}^\top(\m{g}-\m{g}^*)\right)^2}{\|\bm{\delta}\|^2} ~\text{(due to Eqn.~\eqref{eq:20})}\\
    &\geq \frac{1}{4\max_{(i,j)\in \mc{E}} \|\m{z}_{ij}\|^2} \frac{(\m{p}^\top\m{L}_b^*\m{p})^2}{\|\bm{\delta}\|^2} ~\text{(due to Eqn.~\eqref{eq:lem5a})}\\
    &\geq \underbrace{\frac{\lambda_{\min}^2(\m{L}_{ff}^*)}{4\lambda_{\max}(\m{L}_{ff}^*)}}_{:=\chi} \frac{\min_{(i,j)\in \mc{E}}\|\m{z}_{ij}\|}{\max_{(i,j)\in \mc{E}}\|\m{z}_{ij}\|^2} \m{p}^\top\m{r},
\end{align*}
where we have assumed that $\m{g}_{ij}^\top\m{g}_{ij}^*\geq 0,~\forall (i,j)\in \mc{E}$.
Using the inequalities \eqref{eq:26} and \eqref{eq:29}, and let $x=\|\bm{\delta}\|$, we have
\begin{align*}
    f(x) &= \frac{\min_{(i,j)\in \mc{E}}\|\m{z}_{ij}\|}{\max_{(i,j)\in \mc{E}}\|\m{z}_{ij}\|^2} \geq \frac{N-\sqrt{n}x}{(\sqrt{n}x+M)^2},\\
    \frac{\partial f(x)}{\partial x} &= \frac{x-(M+2N)/\sqrt{n}}{n(\sqrt{n}x+M)^3}.
\end{align*}
With $\zeta \geq 0$ such that $g(\zeta) = \frac{\zeta+ \sqrt{{n}\zeta^2 + 4{M}\zeta}}{2} < \frac{N}{\sqrt{n}}$, by similar arguments as in Lemma~\ref{lem:4}, we have $0\leq x\leq g(\zeta)<\frac{N}{\sqrt{n}}$. Thus,  $\frac{\partial f(x)}{dx}<0$ and $f(x) \geq f(g(\zeta))>0$. Notice that $g(\zeta)$ is strictly increasing in $\zeta$.\\
Now, let $V(0)<\vartheta \frac{\zeta}{K}$ with $K>1$, we have $\|\bm{\delta}(0)\|\leq g(\zeta/K) < g(\zeta)$, and thus if $\bm{\delta}(t)$ increases, there must be a time interval such that $\bm{\delta}(t) \in [g(\zeta/K),g(\zeta)]$. During such a time in interval,  
\begin{align*}
    \dot{V} &\leq - k_p\chi_1 f(g(\zeta)) \m{p}^\top \m{r} - \alpha\sum_{i=l+1}^n\frac{(\gamma_i - \gamma_0)^2}{2} + \frac{n-l}{2}\alpha\gamma_0^2\\
    &\leq - \underbrace{\min\{k_p\chi f(g(\zeta)),\alpha\kappa\}}_{:=\varrho} V + \frac{n-l}{2}\alpha\gamma_0^2\\
    &\leq -\varrho(1-\theta) V - \varrho\theta V + \frac{n-l}{2}\alpha\gamma_0^2,
\end{align*}
for $\theta \in (0,1)$. If 
\begin{align} \label{eq:33}
    g\left(\frac{n-l}{2\vartheta\varrho\theta} \alpha\gamma_0^2\right)<g(\zeta)<\frac{N}{\sqrt{n}},
\end{align}
for
\[\vartheta\zeta > V(t) \geq \frac{n-l}{2\varrho\theta} \alpha\gamma_0^2, \]
we still have $\|\bm{\delta}\| < g(\zeta)$, and thus the inequality $\dot{V}\leq -\varrho(1-\theta) V$ holds. This implies that $V$ will decrease, and thus $\|\bm{\delta}\| \in [0,g(\zeta))$,~$\forall t\geq 0$. Finally, the inequality \eqref{eq:33} is always feasible given that $\zeta$ is selected sufficiently small and $k_p\chi$, $\alpha\kappa$ are selected large enough.
\end{Remark}}

\section{Application in formation tracking}
Let the leaders move with the same velocity $\m{v}^*(t)$, which is assumed to be a bounded,  uniformly continuous function. The desired formation $\m{p}^*$ in Definition \ref{def:target_formation} is now time-varying, with $\dot{\m{p}}^* = \m{1}_n \otimes \m{v}^*$. Thus, it is assumed that $(\mc{G},\m{p}^*(0))$ is infinitesimally rigid in $\mb{R}^d$. We will show that the adaptive formation control laws \eqref{eq:bearing_based_control_law} and \eqref{eq:Bearing_OnlyC} are still capable of stabilizing the desired leader-follower formation.

The motion of the $n$-agent system under the control law \eqref{eq:bearing_based_control_law} is now given in matrix form as follows:
\begin{align} \label{eq:moving_system}
\dot{\m{p}} = \begin{bmatrix}
\dot{\m{p}}^{\rm L}\\
\dot{\m{p}}^{\rm F}
\end{bmatrix} = \begin{bmatrix}
\m{0}_{dl}\\
\dot{\m{p}}^{\rm F} - \m{1}_f \otimes \m{v}^*
\end{bmatrix} + \m{1}_n \otimes \m{v}^*.
\end{align}
Let $\boldsymbol{\delta} = \m{p} - \m{p}^* = \begin{bmatrix}
\m{0}_{dl}\\
{\m{p}}^{\rm F} - {\m{p}}^{\rm F*}
\end{bmatrix}$, then 
\begin{align*}
\dot{\boldsymbol{\delta}} = \begin{bmatrix}
\m{0}_{dl}\\
\dot{\m{p}}^{\rm F} - \m{1}_f \otimes \m{v}^*
\end{bmatrix} = \bar{\m{Z}} \left( \begin{bmatrix}
\m{0}_{dl}\\
\dot{\m{p}}^{\rm F}
\end{bmatrix} - \m{1}_n \otimes \m{v}^* \right).
\end{align*} 

Suppose that the displacement-based control law \eqref{eq:bearing_based_control_law} is adopted for followers, we have
\begin{align}
&\dot{\boldsymbol{\delta}} =- \bar{\m{Z}}((\tilde{\m{R}}_b^{*})^\top \bar{\boldsymbol{\Gamma}}{\rm sgn}({\rm blkdiag}(\m{P}_{\m{g}^*_k})\bar{\m{H}}\boldsymbol{\delta}) - \m{d} + \m{1}_n \otimes \m{v}^*),\nonumber\\
&\dot{\boldsymbol{\gamma}} = \kappa \left[\|\m{q}_1\|_1,\ldots,\|\m{q}_m\|_1\right]^\top, \label{eq:bearing_based_maneuver_system}
\end{align}
which is of the same form as \eqref{eq:bearing_based_system}, but having an additional disturbance term $-\bar{\m{Z}}(\m{1}_n \otimes \m{v}^*)$. Thus, the following theorem can be proved.
\begin{Theorem} \label{thm:3}
Consider the $n$-agent system \eqref{eq:bearing_based_maneuver_system} under the displacement-based control law \eqref{eq:bearing_based_control_law}, the following statements hold:
\begin{enumerate}[label = (\roman*)]
    \item $\boldsymbol{\delta}(t) \to \m{0}_{dn}$, as $t \to + \infty$, 
    \item There exists a constant vector $\boldsymbol{\gamma}^*=[\ldots,\gamma_{ij}^*,\ldots]^\top = [\gamma_1^*,\ldots,\gamma_m^*]^\top$, such that $\boldsymbol{\gamma}(t) \to \boldsymbol{\gamma}^*$, as $t \to + \infty$,
    \item Additionally, if $\gamma_k^*> \gamma_0':=(\beta + \|\m{v}^*\|_{\infty})\sqrt{\frac{dn}{\lambda_{\min}(\m{L}_{ff}^*)}},~\forall k=1,\ldots,m$, and there exists a finite time $T$ such that $|\gamma_k - \gamma_k^*|< \min_k|\gamma_k - \gamma_0|,~\forall i=1,\ldots,m,$ then $\boldsymbol{\delta}(t) \to \m{0}_{dn}$ in finite time.
\end{enumerate}
\end{Theorem}
\begin{IEEEproof}
The proof is similar to the proof of Theorem \ref{thm:1} and will be omitted.
\end{IEEEproof}

Finally, if the bearing-only control law \eqref{eq:Bearing_OnlyC} is adopted for followers, the $n$-agent formation can be expressed in matrix form as 
\begin{equation}
\label{eq:bearing_only_maneuver_system2}
\begin{split}
    \dot{\boldsymbol{\delta}} &= \bar{\m{Z}} \left(-\bar{\boldsymbol{\Gamma}}  \text{sgn}\left( \bar{\m{H}}^\top (\m{g}- \m{g}^*) \right) + \m{d}-\m{1}_n \otimes \m{v}^*\right), \\
    \dot{\boldsymbol{\gamma}} &= \kappa \left[\ldots,\left|\left|\sum_{j\in\mc{N}_i}(\m{g}_{ij} - \m{g}_{ij}^*)\right|\right|_1,\ldots \right]^\top, 
\end{split}
\end{equation}
which is of the same form as \eqref{eq:system_BOM}--\eqref{eq:system_BOM1}, but having an additional unknown disturbance term $-\bar{\m{Z}}(\m{1}_n \otimes \m{v}^*)$. We have the following theorem, whose proof is similar to the proof of Theorem~\ref{thm:2} and will be omitted.
\begin{Theorem} \label{thm:4}
Consider the $n$-agent system \eqref{eq:bearing_only_maneuver_system2} under the adaptive bearing-only based control law \eqref{eq:Bearing_OnlyC}. There exists a positive constant $\alpha>0$ such that if the Lyapunov function in \eqref{eq:Lyap_bom} satisfies $V(0)<\alpha$, {there will be no collision between agents in formation,} $\lim_{t\to +\infty}(\m{p}(t)-\m{p}^*) = \m{0}_{dn}$, and $\lim_{t\to +\infty}\boldsymbol{\gamma}(t) = \boldsymbol{\gamma}^*$, for some constant vector $\boldsymbol{\gamma}^*$.
\end{Theorem}

\begin{Remark} In formation tracking, the leaders' trajectories can be embedded into each leader from the beginning, or can be remotely regulated by a control center. The leader agents are assumed to be equipped with a better positioning system, so that their positions are available for control and monitoring objectives. Suppose that the leaders are also subjected to bounded unknown disturbances, i.e.,
\begin{align*}
\dot{\m{p}}_i(t) &= \m{u}_i(t) + \m{d}_i(t),~\forall i = 1, \ldots, l,
\end{align*}
where $\|\m{d}_i\| < \beta$. To ensure that the leaders track their desired trajectories $\m{p}_i^*(t)$, and thus, eventually act as moving references for follower agents, the following position tracking law is respectively proposed
\begin{align*}
\m{u}_i(t) = -k_p(\m{p}_i - \m{p}_i^*) - \beta_1 {\rm sgn}(\m{p}_i - \m{p}_i^*),
\end{align*}
where $\beta_1 > \beta$. By considering the Lyapunov function $V = \frac{1}{2}\|\m{p}_i - \m{p}_i^*\|^2$, we can prove that $\m{p}_i(t) \to \m{p}_i^*$ in finite time.
\end{Remark}

\begin{Remark} In this remark, we discuss the implementation of the bearing-only formation control laws. For indoor (laboratory) environments, the bearing vectors can be obtained from a single omnidirection camera mounted on the agent. Another setup is using an indoor localization system, which localizes agents' positions, calculates the bearing vectors, and sends this information to each agent to determine a corresponding control input \cite{Ko2020bearing,zhao2019bearing}. For outdoor implementation, the authors in \cite{Schilling2021vision} proposed to use four cameras attached to four sides of a quadcopter to  obtain bearing vector information from different directions. The limited field-of-view of a camera can be considered in the control law, as proposed in \cite{fabris2022bearing}.
\end{Remark}

\section{Simulation results}
\label{sec:5}
In this section, we provide a few {simulations} to demonstrate the effectiveness of the formation control laws proposed in Sections \ref{sec:3}, \ref{sec:4}, and \ref{sec:5}. In all simulations, the target formation is described by a graph $\mc{G}$ of 20 vertices and 39 edges and a desired configuration $\m{p}^*$ (a dodecahedron) as depicted in Figure~\ref{fig:desiredFormation}. It can be checked that $(\mc{G},\m{p}^*)$ is infinitesimally bearing rigid in 3D. In the simulations, there are $l=3$ leaders and $17$ followers. 

\begin{figure}[t]
\centering
\begin{subfloat}[]{
\includegraphics[width=0.45\textwidth]{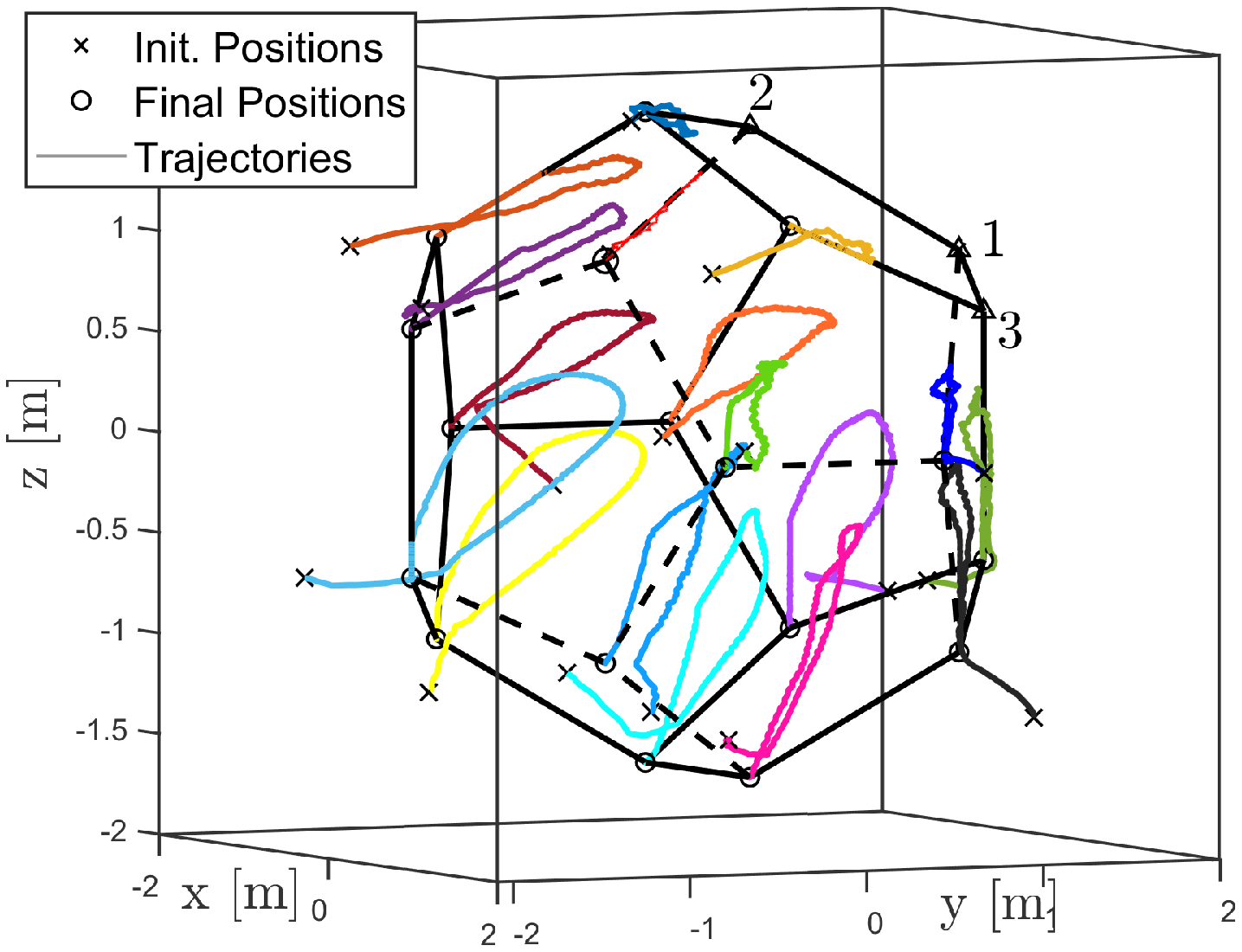}
\label{fig:sim1atraj1}}
\end{subfloat}\hfill
\begin{subfloat}[]{
\includegraphics[width=0.45\textwidth]{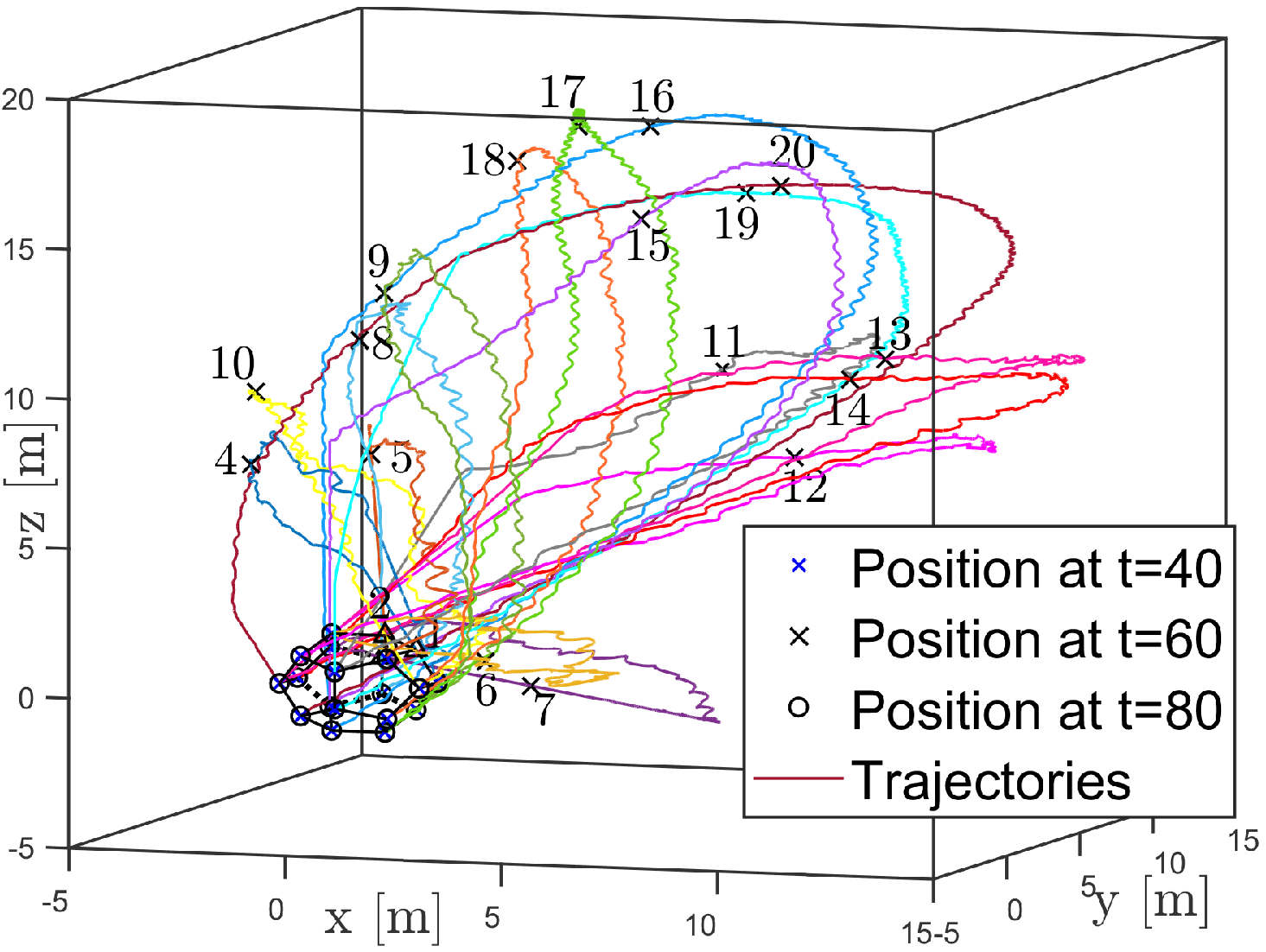}
\label{fig:sim1atraj2}}
\end{subfloat}\hfill
\begin{subfloat}[]{
\includegraphics[width=0.23\textwidth]{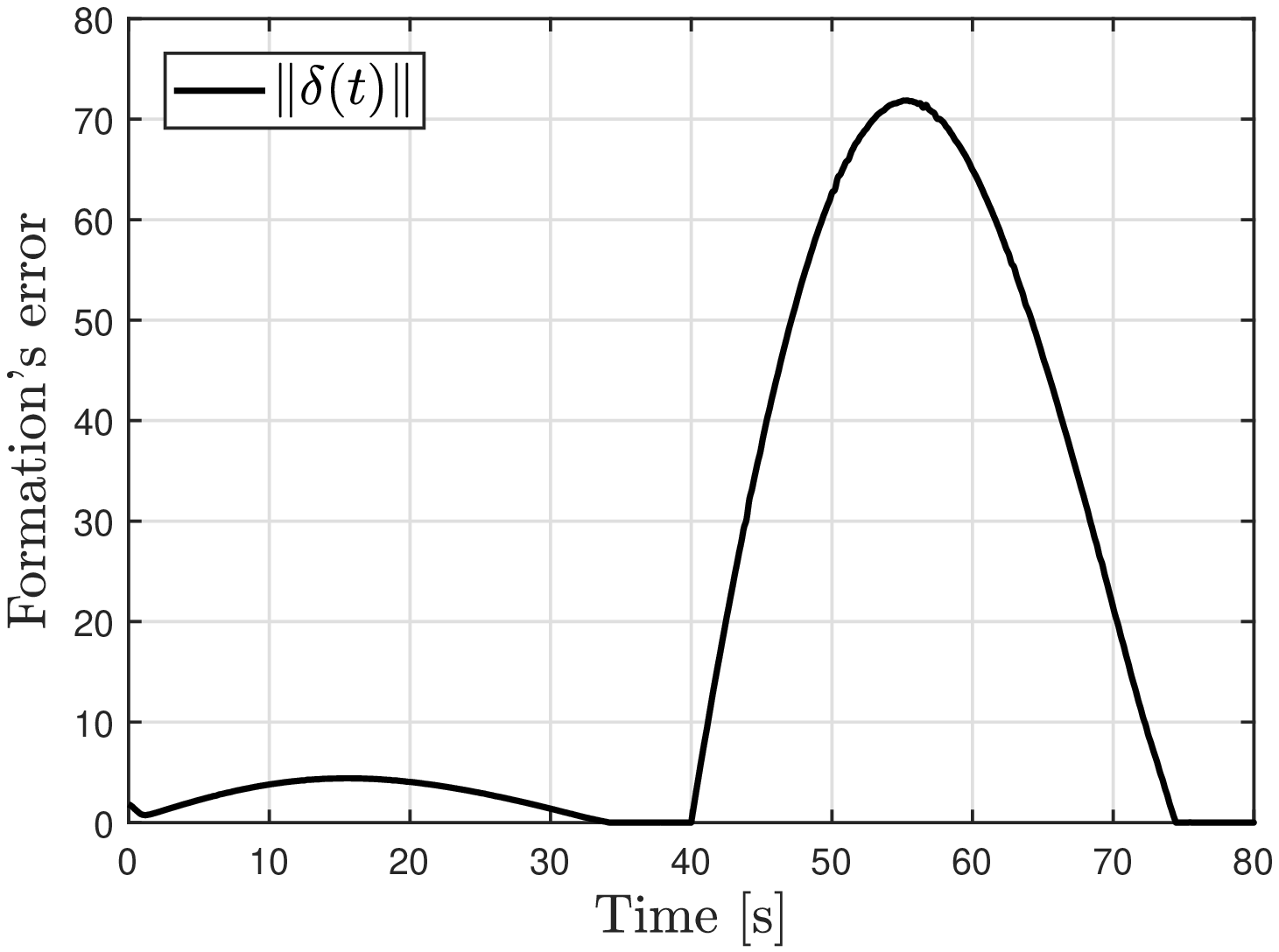}
\label{fig:sim1aerr}}
\end{subfloat}\hfill
\begin{subfloat}[]{
\includegraphics[width=0.23\textwidth]{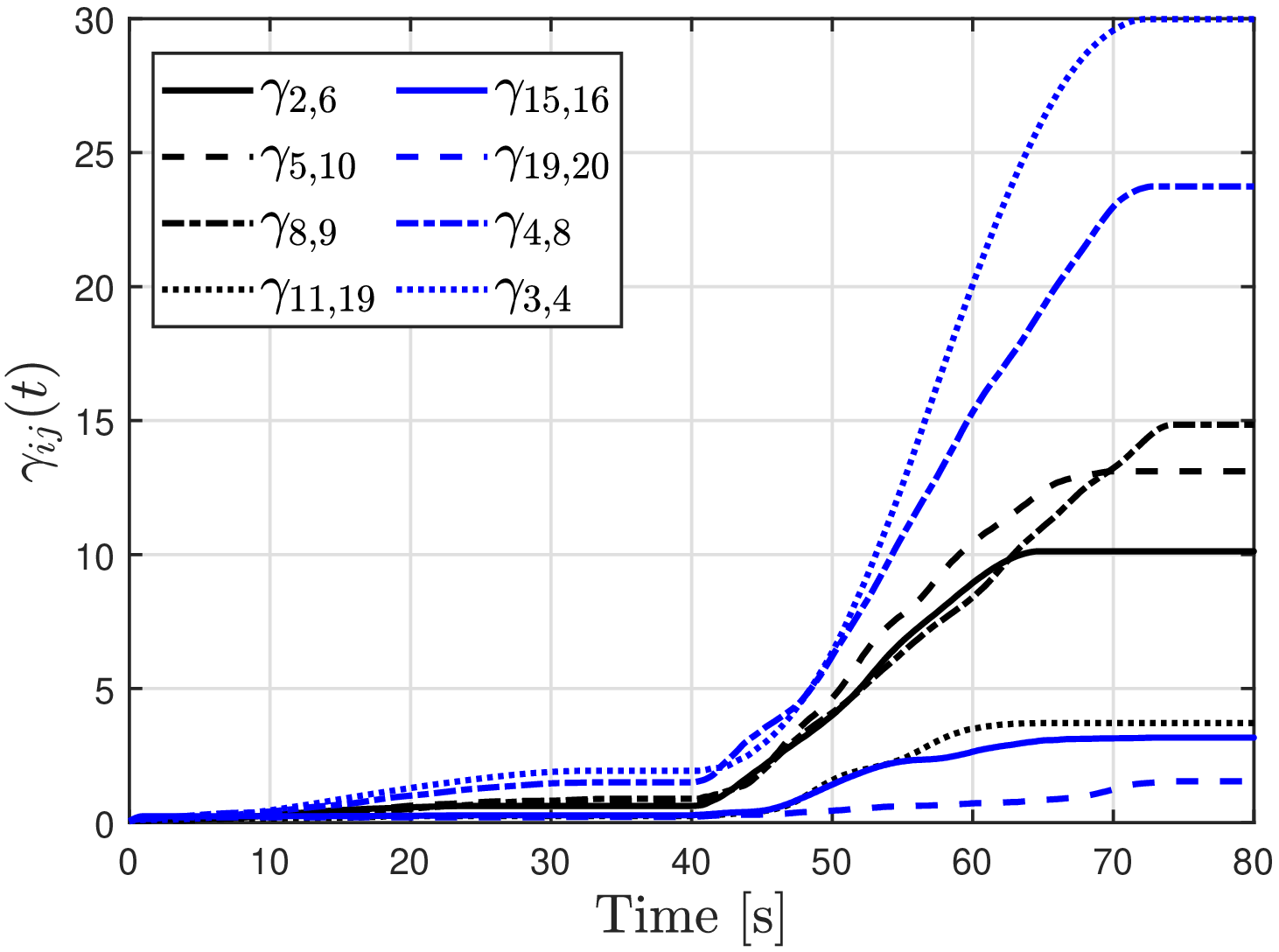}
\label{fig:sim1agamma}}
\end{subfloat}
\caption{Simulation 1a: the 20-agent system under the control law \eqref{eq:bearing_based_control_law}. (a) Trajectories of agents from 0 to 40 seconds (leaders are marked with $\Delta$, followers' initial and final positions ($t=40$ sec) are marked with `${\rm x}$' and `${\rm o}$', respectively); (b) Trajectories of agents from 40 to 80 seconds; (c) Formation's error versus time; (d) A subset of the adaptive gains $\gamma_{ij}$ versus time. \label{fig:Sim1a}}
\end{figure}

\begin{figure}[ht]
\centering
\begin{subfloat}[]{
\includegraphics[width=0.33\textwidth]{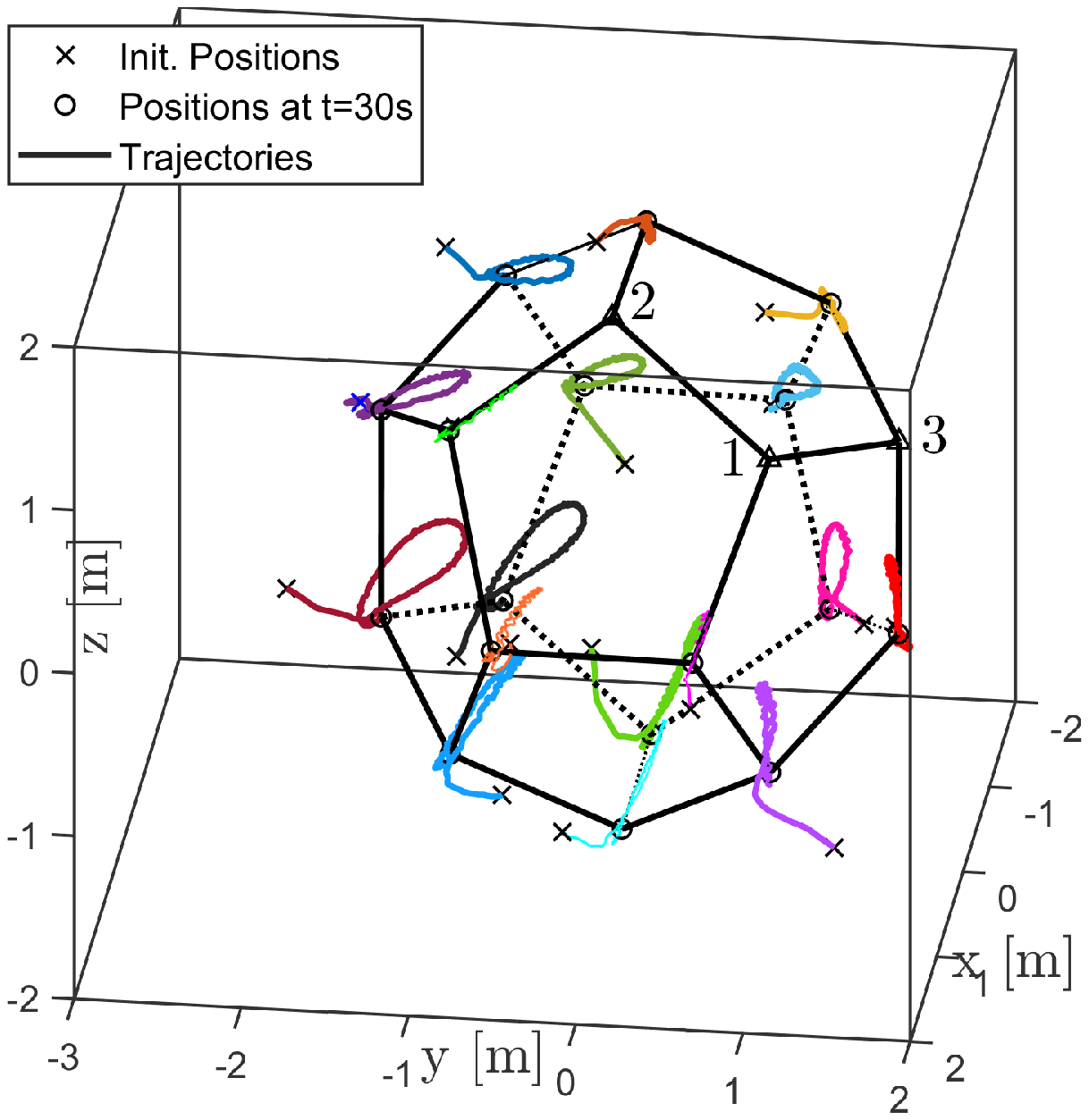}
\label{fig:sim1btraj1}}
\end{subfloat}\hfill
\begin{subfloat}[]{
\includegraphics[width=0.35\textwidth]{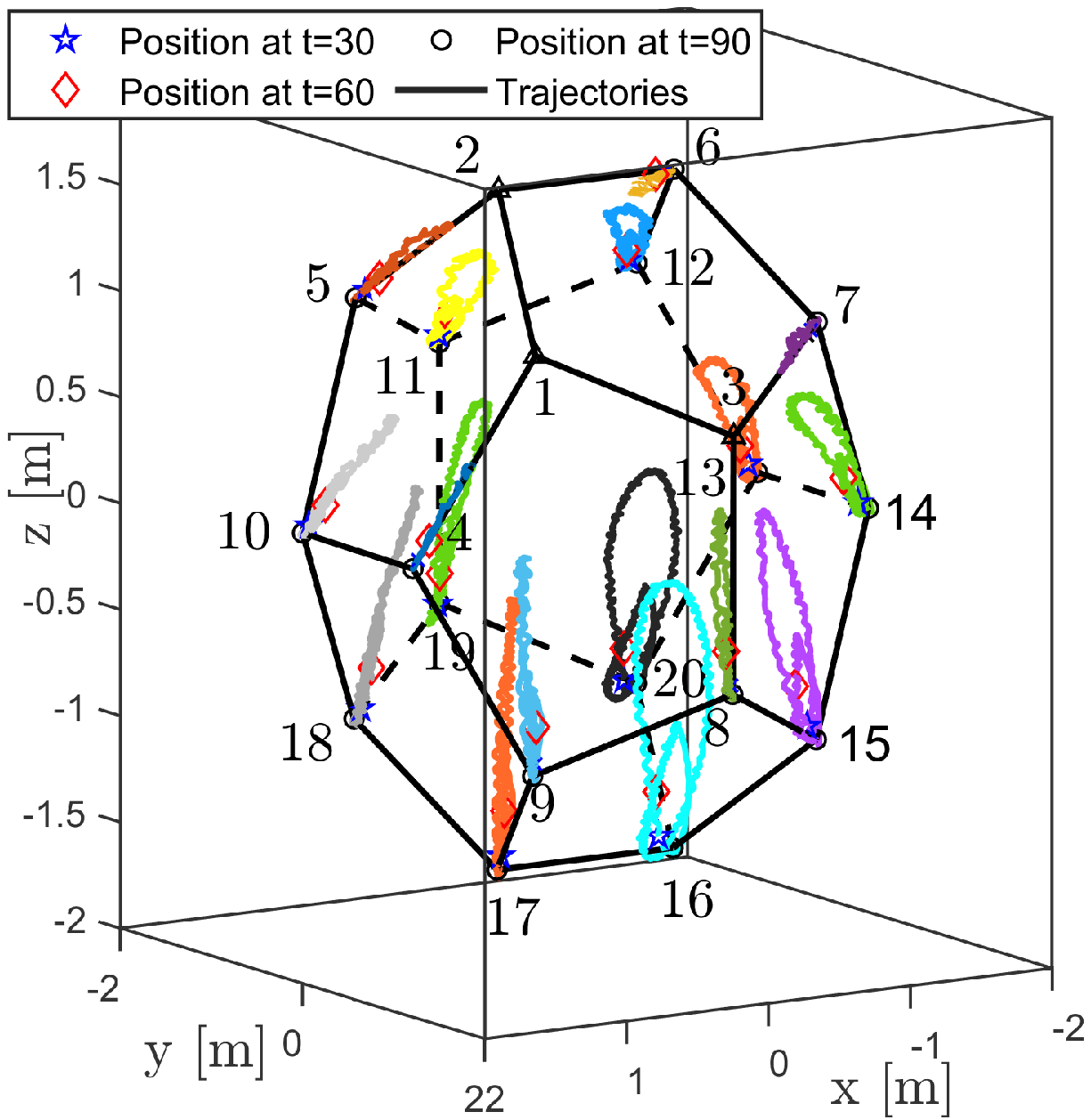}
\label{fig:sim1btraj2}}
\end{subfloat}\hfill
\begin{subfloat}[]{
\includegraphics[width=0.23\textwidth]{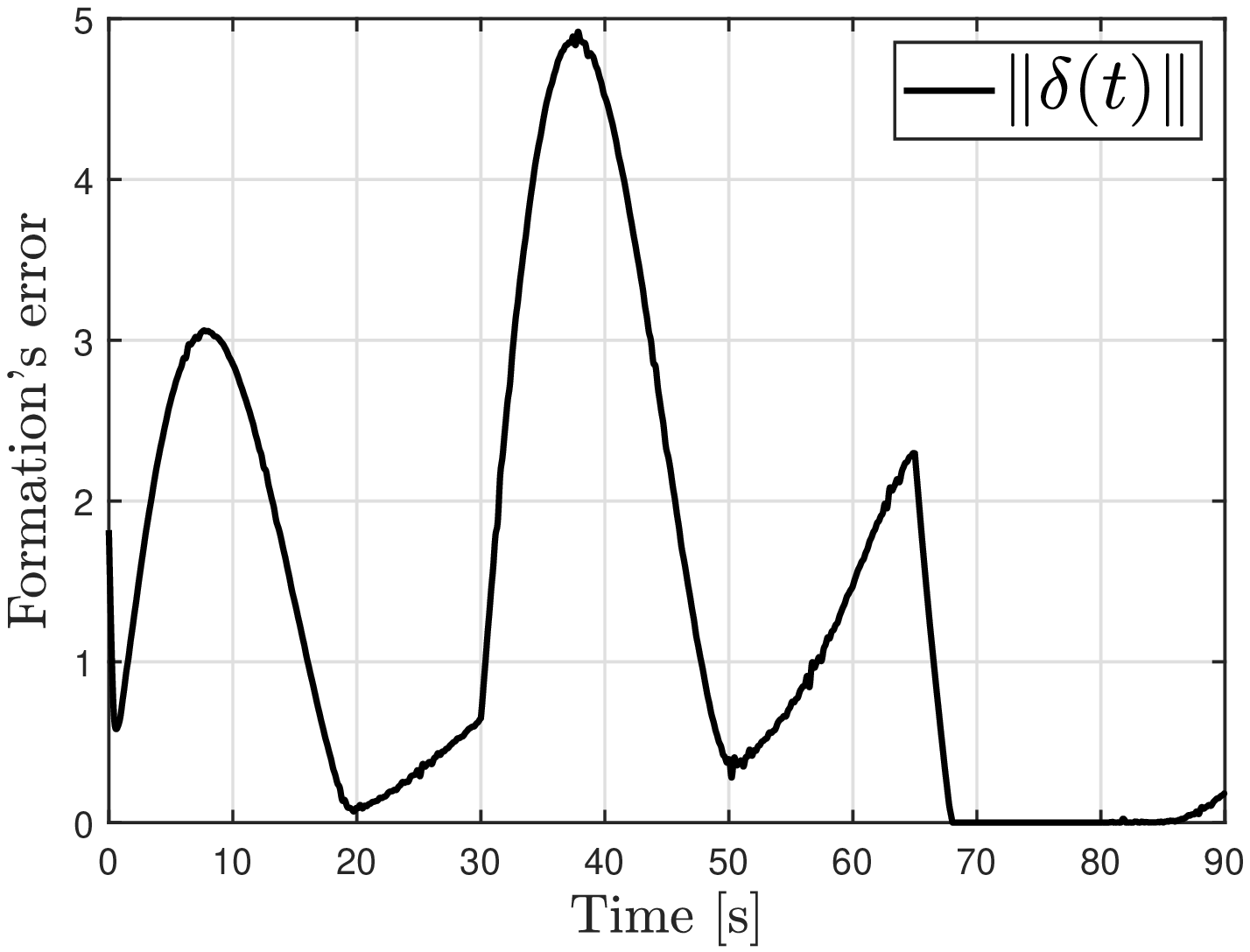}
\label{fig:sim1berr}}
\end{subfloat}\hfill
\begin{subfloat}[]{
\includegraphics[width=0.23\textwidth]{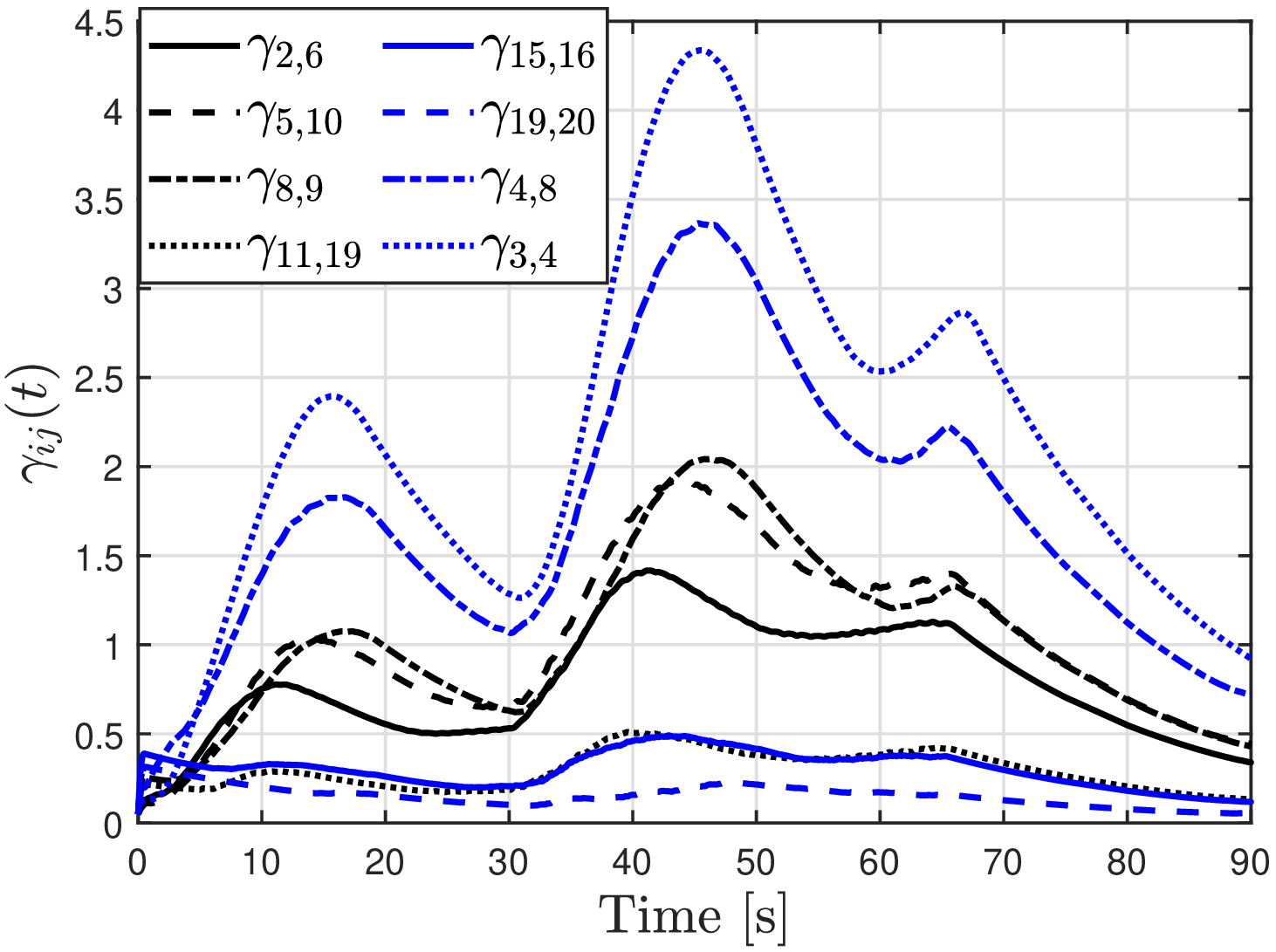}
\label{fig:sim1bgamma}}
\end{subfloat}
\caption{Simulation 1b: the 20-agent system under the control law \eqref{eq:bearing_based_control_law2}. (a) Trajectories of agents from 0 to 30 seconds (leaders are marked with $\Delta$, followers' initial and final positions are marked with `${\rm x}$' and `${\rm o}$', respectively); (b) Trajectories of agents from 30 to 90 seconds; (c) Formation's error versus time; (d) A subset of the adaptive gains $\gamma_{ij}$ versus time. \label{fig:Sim1b}}
\end{figure}
\subsection{Bearing-based formation control with disturbance rejection}
First, we simulate the formation with the control law \eqref{eq:bearing_based_control_law}. 
Let each follower $i$ be modeled by a single integrator with disturbance $\m{d}_i$ given as 
\begin{align}
\m{d}_i(t) = \begin{cases}
0.1 \m{h}_i(t), & \mbox{if } 0\leq t \leq 40 s, \\
0.15 \m{h}_i(t), & \mbox{if } t \geq 40 s,
\end{cases}
\end{align}
where $\m{h}_i(t) = [\sin(it)+1,~\cos(it)+\tanh(t),~1-e^{-it}]^\top$.

The control law \eqref{eq:bearing_based_control_law} is used with $\kappa=0.2$ and $\gamma_i(0),~i=4,\ldots, 20,$ are randomly generated on the interval $[0,0.05]$. Simulation results are given as in Fig.~\ref{fig:Sim1a}.

According to Figs.~\ref{fig:sim1atraj1}, \ref{fig:sim1aerr}, and \ref{fig:sim1agamma}, for $0\leq t \leq 40$ seconds, the desired formation is asymptotically achieved and the adaptive gains $\gamma_{ij}$ increase until the corresponding bearing constraint is stabilized. From $t=40$ seconds, the magnitude of the disturbance suddenly increases, which drives the agents out of the desired formation. The errors invoke the adaptive mechanism, $\gamma_{ij}$ increase again. It can be seen from Figs.~\ref{fig:sim1atraj2}, \ref{fig:sim1aerr}, and \ref{fig:sim1agamma} that followers are driven out from their desired positions from 40 to 55 seconds, as  the magnitudes of their formation control laws are not big enough to counter the disturbance. From 55 to 80 seconds, when $\gamma_{ij}$ are sufficiently large, the agents are pulling back to the desired positions, and the desired formation is eventually achieved. 

Second, we conduct a simulation of the formation under the adaptive control law with increasing/decreasing gains \eqref{eq:bearing_based_control_law2}. The disturbance acting on a follower $i$ in this simulation is given as 
\begin{align}
\m{d}_i(t) = \begin{cases}
0.15 \m{h}_i(t), & \mbox{if } 0\leq t \leq 30 s, \\
0.3 \m{h}_i(t), & \mbox{if } 30 \leq t \leq 65 s, \\
0.1 \m{h}_i(t), & \mbox{if } 65 \leq t \leq 90 s.
\end{cases}
\end{align}
With $\gamma_{ij}(0)=0.05, \forall (i,j) \in \mc{E}$, $k_p = 0.5$ (proportional gain), $\kappa=1$ (rate of adaptation), $\alpha = 0.05$ (leakage coefficient), and $\m{p}(0)$ being chosen the same as the previous simulation, we obtain the simulation results as depicted in Fig.~\ref{fig:Sim1b}.

As shown in Figs.~\ref{fig:sim1btraj1}, \ref{fig:sim1berr}, and \ref{fig:sim1bgamma}, for $0\leq t \leq 15$ seconds, the adaptive gains $\gamma_{ij}$ increase and the control law drives the agents to a neighborhood of the desired formation. Due to the existence of a leakage term {$-\kappa\alpha \gamma_{ij}$} in \eqref{eq:bearing_based_control_law2}(c), once a desired bearing constraint is sufficiently small, $\gamma_{ij}$ tends to reduce their values from 15 to 30 seconds. The decrements of $\gamma_{ij}$ make the formation errors raise again, however, $\m{p}(t)$ remains on a small ball $\mc{B}_{R_1}(\m{p}^*)$ centered at $\m{p}^*$, whose radius $r$ is jointly determined by the controller's parameters, the desired formation, and the magnitude of the unknown disturbance.

From $t=30$ to $45$ seconds, as the magnitude of the disturbance is doubled, the agents are out from $\mc{B}_{R_1}(\m{p}^*)$. As the errors increase, the term $\|\m{q}_{ij}\|$ dominates the leakage term in the adaptive mechanism \eqref{eq:bearing_based_control_law2}(c), and thus $\gamma_{ij}$ increase again. It can be seen from Figs.~\ref{fig:sim1btraj2}, \ref{fig:sim1berr}, and \ref{fig:sim1bgamma} that followers are driven further from their desired positions from 30 to about 38 seconds, and then being attracted to a ball $\mc{B}_{R_2}(\m{p}^*)$ centered at $\m{p}^*$, with $R_2>R_1$, from 38 to 65 seconds. For $45 \leq t \leq 65$, the bearing constraints are sufficiently small, it can be seen that $\gamma_{ij}$ decrease again due to the leakage term. For $t\geq 65$, as the disturbance magnitude decreased to 0.1, as $\gamma_{ij}(t=65 s)$ satisfy the requirement of Lemma \ref{lem:3.2}, $\m{p}$ converges to $\m{p}^*$ after a short time ($\m{p}(t)=\m{p}^*$ at $t=68$ s). However, from $t=68$s, because the leakage term is the only active term in \eqref{eq:bearing_based_control_law2}(c), $\gamma_{ij}$ decreases. Gradually, once the control law cannot fully reject the disturbance, the disturbances make $\m{p}$ out of $\m{p}^*$. The control law will still keep $\m{p}$ inside a ball $\mc{B}_{R_3}(\m{p}^*)$ centered at $\m{p}^*$, with $R_3<R_1$. 

\subsection{Bearing-only formation control with disturbance rejection}
In this subsection, we simulate the adaptive bearing-only control law \eqref{eq:Bearing_OnlyC} for the 20-agent system. The simulation's parameters are $\kappa_i = 2$, and $\gamma_{ij}(0) = 0.5$. 
\begin{figure*}
\begin{subfloat}[]{
\includegraphics[width=0.31\textwidth]{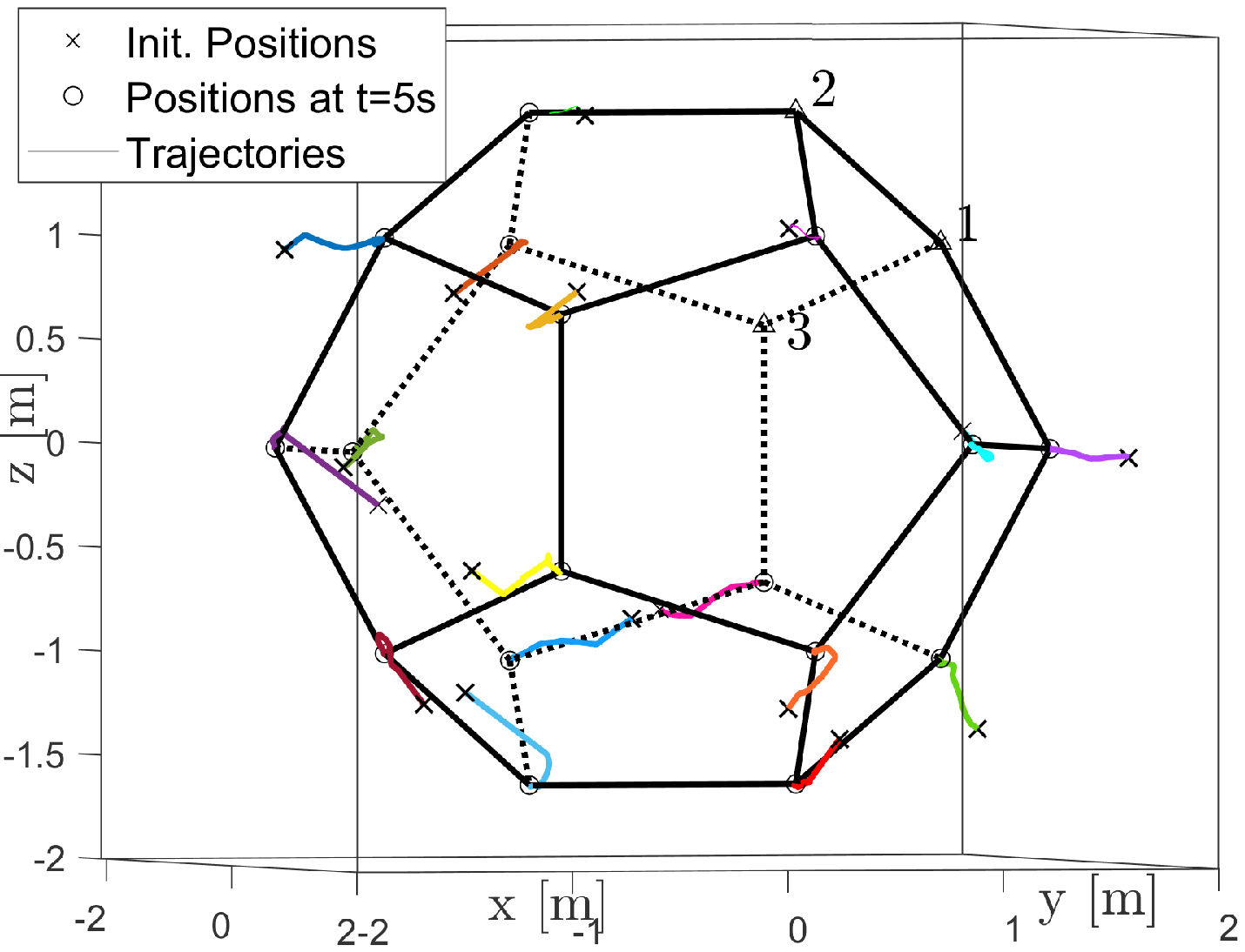}
\label{fig:sim2traj1}}
\end{subfloat}
\hfill
\begin{subfloat}[]{
\includegraphics[width=0.31\textwidth]{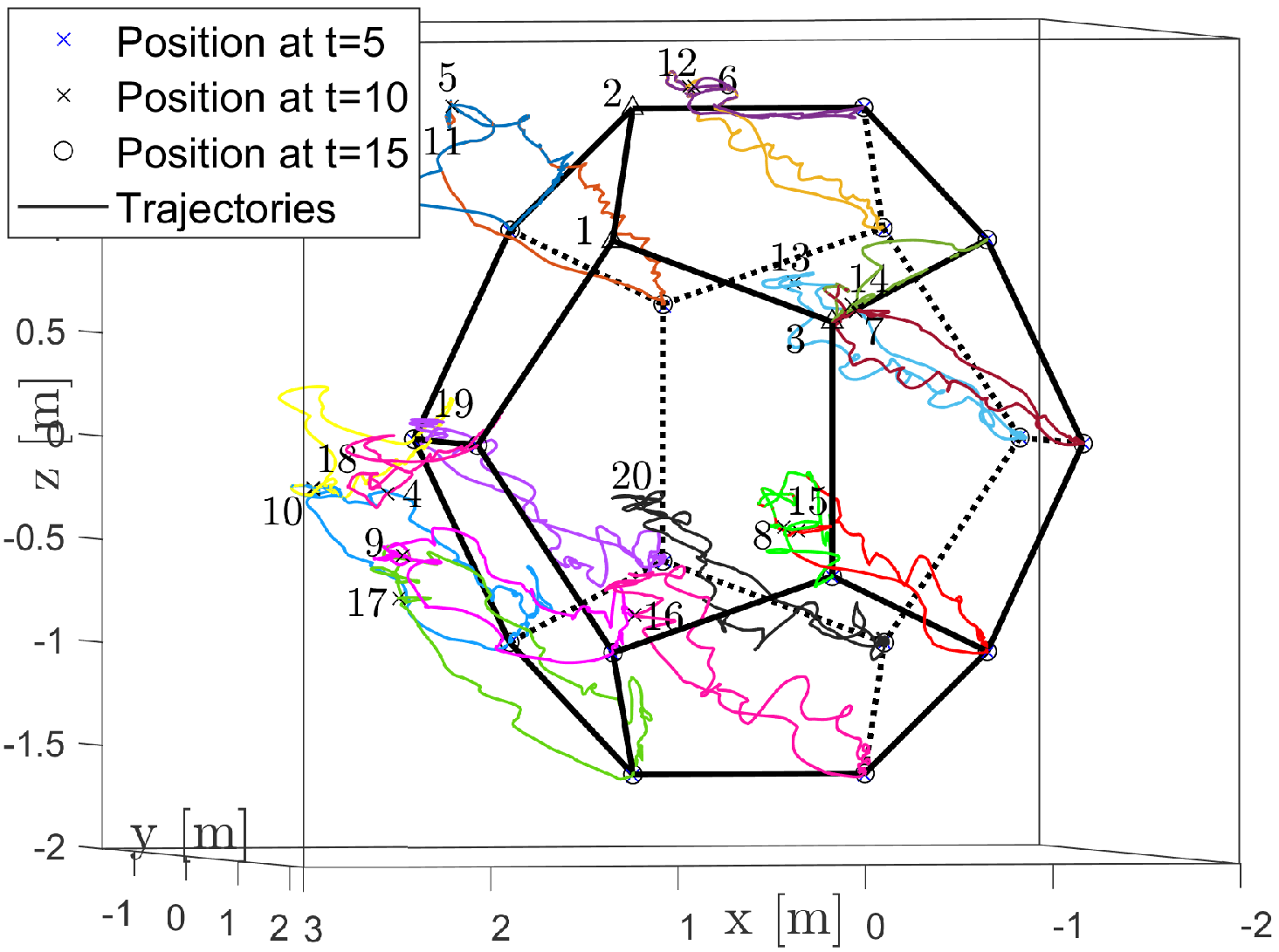}
\label{fig:sim2traj2}}
\end{subfloat}
\hfill
\begin{subfloat}[]{
\includegraphics[width=0.31\textwidth]{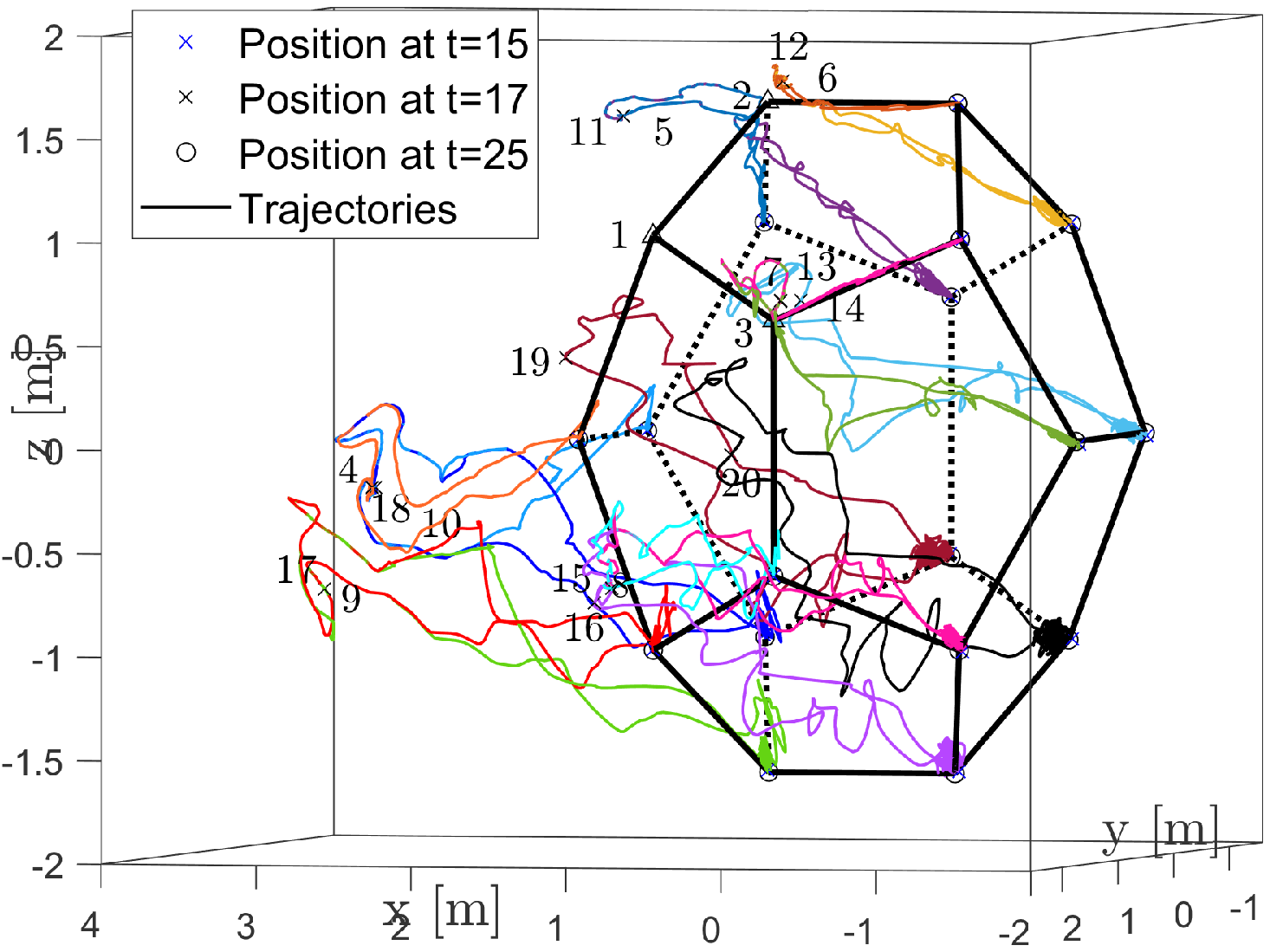}
\label{fig:sim2traj3}}
\end{subfloat}
\hfill
\begin{subfloat}[]{
\includegraphics[width=0.31\textwidth]{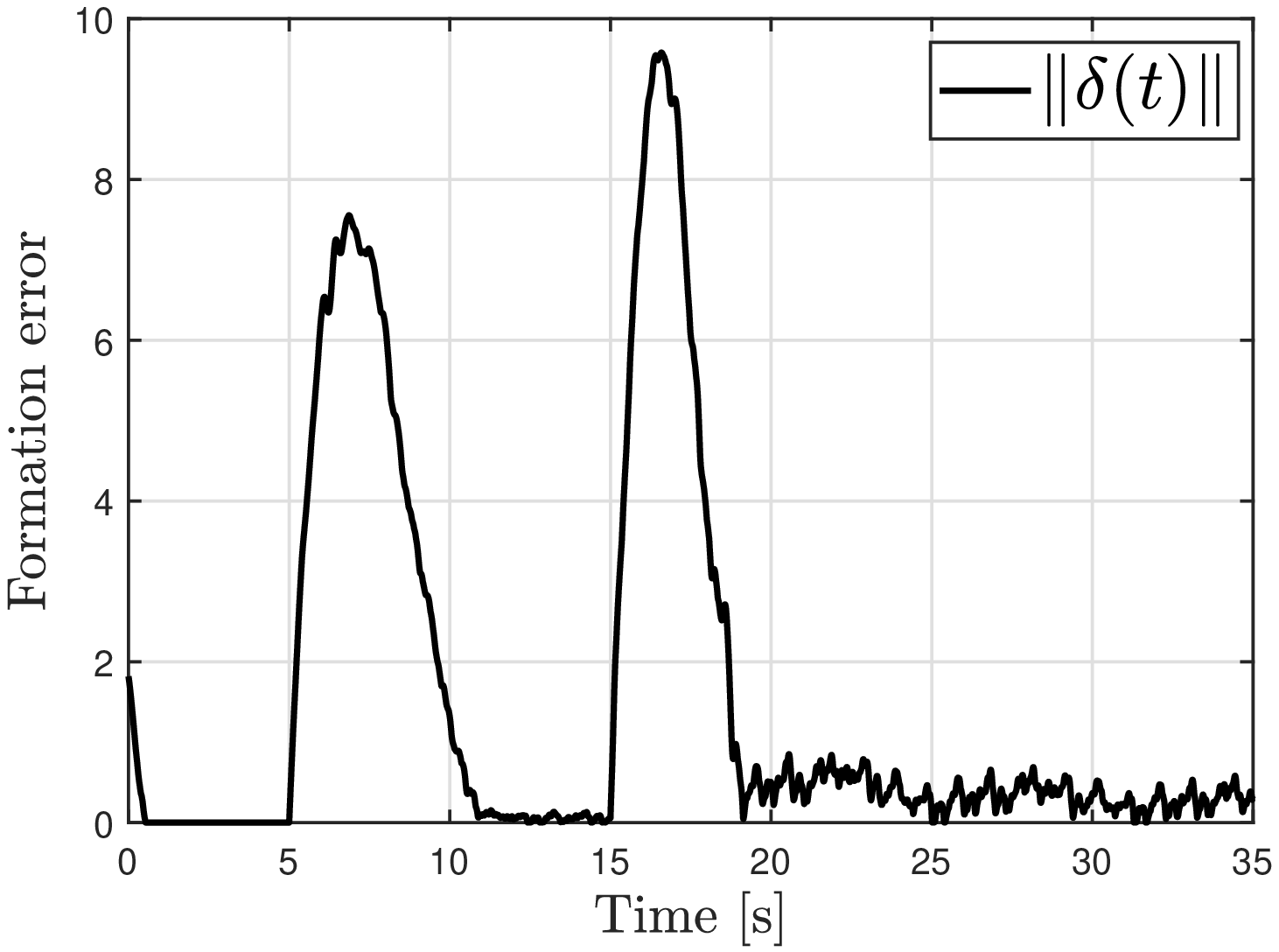}
\label{fig:sim2err}}
\end{subfloat}
\hfill
\begin{subfloat}[]{
\includegraphics[width=0.31\textwidth]{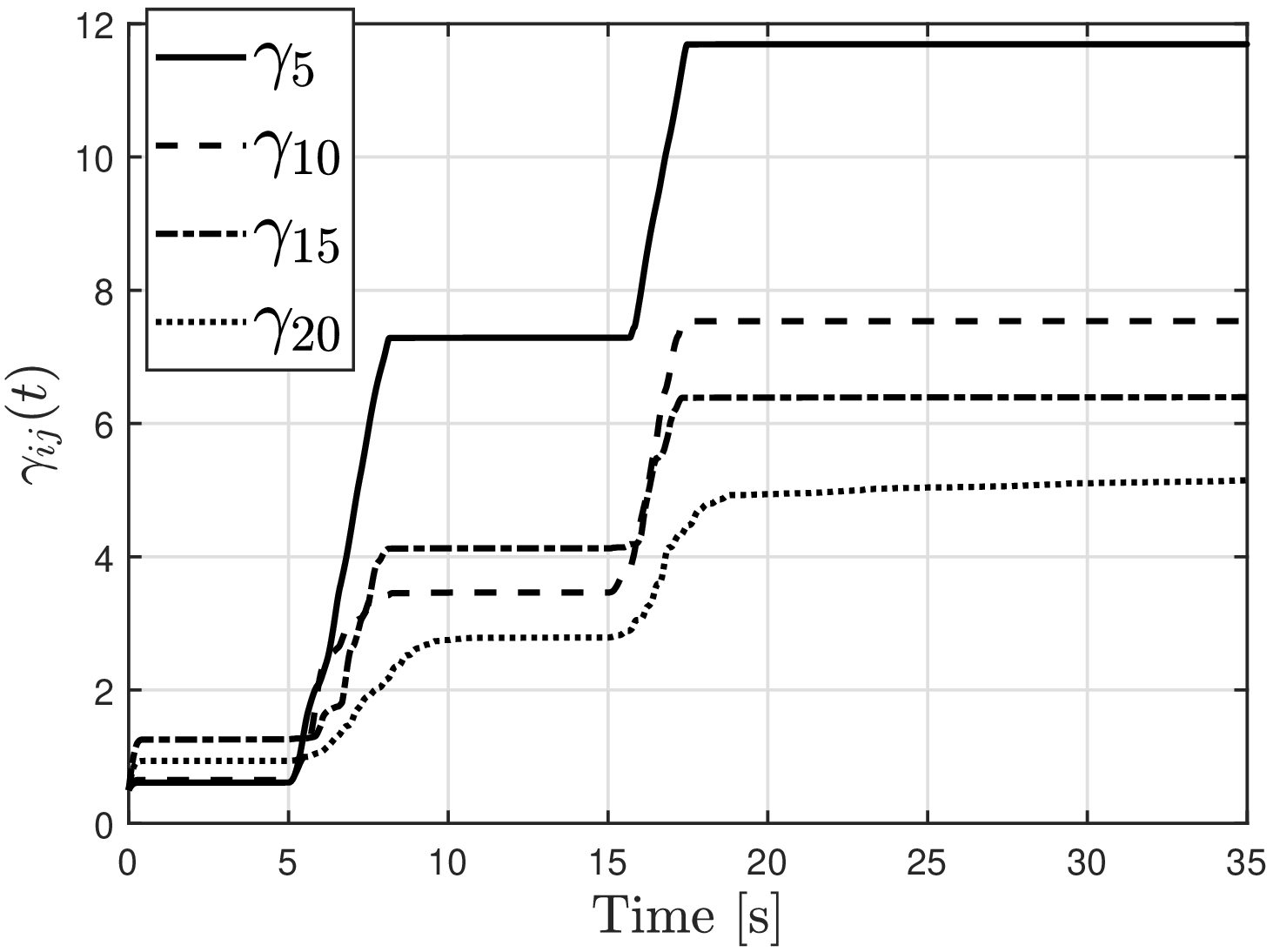}
\label{fig:sim2gamma}}
\end{subfloat}
\hfill
\begin{subfloat}[]{
\includegraphics[width=0.31\textwidth]{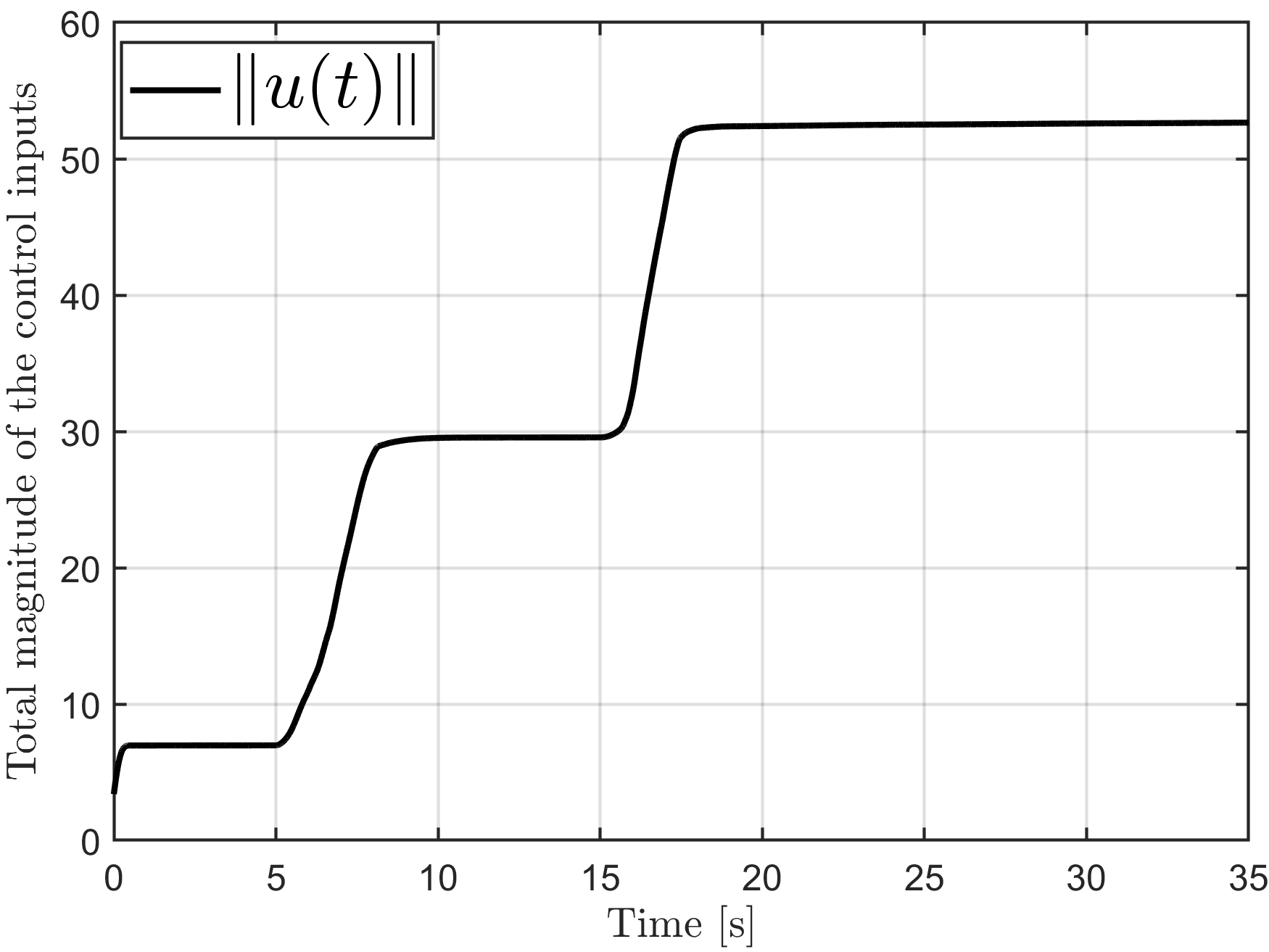}
\label{fig:sim2Input}}
\end{subfloat}
\caption{Simulation 2: the 20-agent system under the bearing-only control law \eqref{eq:Bearing_OnlyC}. (a) Trajectories of agents from 0 to 5 seconds (leaders are marked with $\Delta$, followers' initial and final positions are marked with `${\rm x}$' and `${\rm o}$', respectively); (b) Trajectories of agents from 5 to 15 seconds; (c) Trajectories of agents from 15 to 25 seconds; (d) Formation's error versus time; (e) A subset of the adaptive gains $\gamma_{i}$ versus time. (f) Magnitude of control input versus time. \label{fig:Sim2}}
\end{figure*}
The disturbance of each follower $i$ in this simulation is given as 
\begin{align}
\m{d}_i(t) = \begin{cases}
\m{0}_3, & \mbox{if } 0\leq t \leq 5 s, \\
1.5 \m{h}_i(t), & \mbox{if } 5\leq t \leq 15 s,\\
3 \m{h}_i(t), & \mbox{if } 15 \leq t \leq 25 s.
\end{cases}
\end{align}
The simulation results are depicted in Fig.~\eqref{fig:Sim2}. For $t\in [0,5]$  (second), there is no disturbance acting on the formation, the control law stabilizes $\m{p}(t)$ to $\m{p}^*$ after about 2 seconds. The adaptive gains $\gamma_i$ increase correspondingly in $0\le t \le 2$ and remain unchanging until $t=5s$, when there are disturbances acting on the agents. Due to the presence of the disturbances, $\m{p}$ leaves the target configuration $\m{p}^*$, the bearing errors make $\gamma_i$ increase. In turn, the control law's magnitude increases and is eventually capable of suppressing the disturbance from  $7$s. For $t \geq 7$s, $\m{p}$ approaches to $\m{p}^*$. Approximately, $\m{p}$ reached to $\m{p}^*$ after $13$ seconds, and $\gamma_i$ cease to increase as the bearing constraints were almost satisfied. For $t\geq 15$s, as the disturbances increase their magnitudes, $\m{p}$ leaves $\m{p}^*$ again. The adaptive gains $\gamma_i$ increase correspondingly, and eventually pull $\m{p}$ back to $\m{p}^*$. It can be seen that the increment of $\gamma_{20}$ is relatively slower than other displayed adaptive gains for $20\leq t \leq 35$s. Chattering phenomenon can also be seen due to the disturbances (for $11\leq t \leq 15$ and $20\leq t \leq 35$s), which causes significant fluctuations of $\m{p}$ around $\m{p}^*$.

\subsection{Bearing-based formation tracking}
\label{subsec:6.3}
In this subsection, we simulate the formation  \eqref{eq:bearing_based_maneuver_system} with moving leaders. The leaders' velocities are chosen as 
\[\m{v}^* = \left[\sin\left(\frac{t}{2}\right),~1,~0 \right]^\top,\, t\geq 0.\]
The simulation's parameters are $\kappa = 2$, $\gamma_i(0)=1$. The initial positions of the agents are the same as in the previous simulation. Disturbances are not included in the simulation.

\begin{figure*}
\centering
\begin{subfloat}[]{
\includegraphics[width=.98\textwidth]{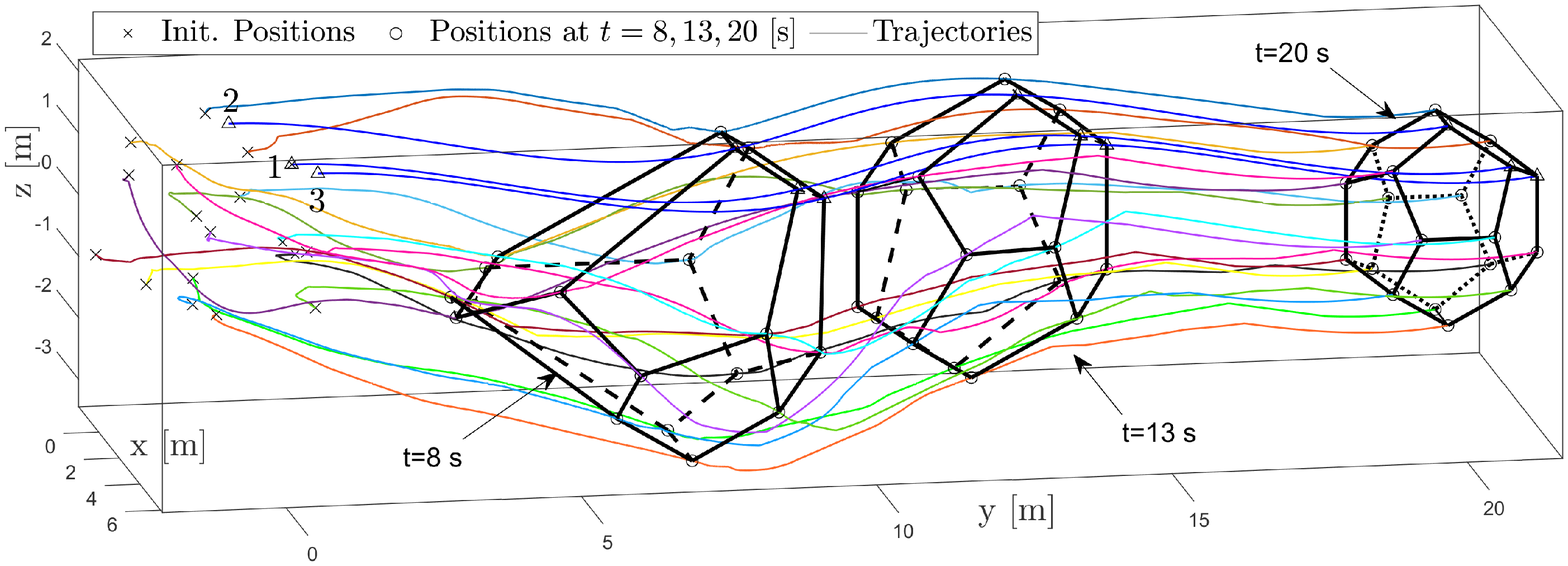}
\label{fig:sim3traj1}}
\end{subfloat}\\
\begin{subfloat}[]{
\includegraphics[width=0.3\textwidth]{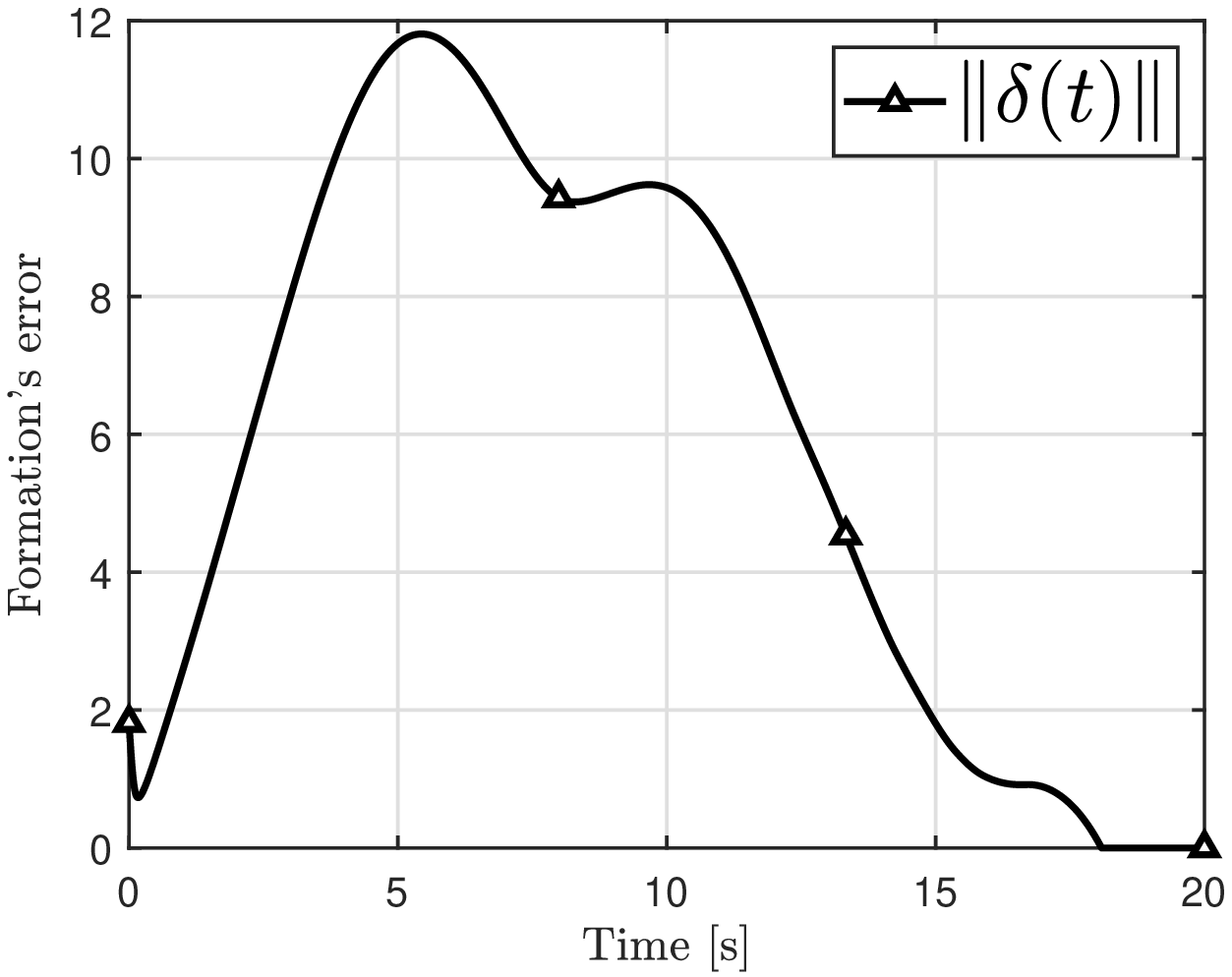}
\label{fig:sim3err}}
\end{subfloat}\hfill
\begin{subfloat}[]{
\includegraphics[width=0.32\textwidth]{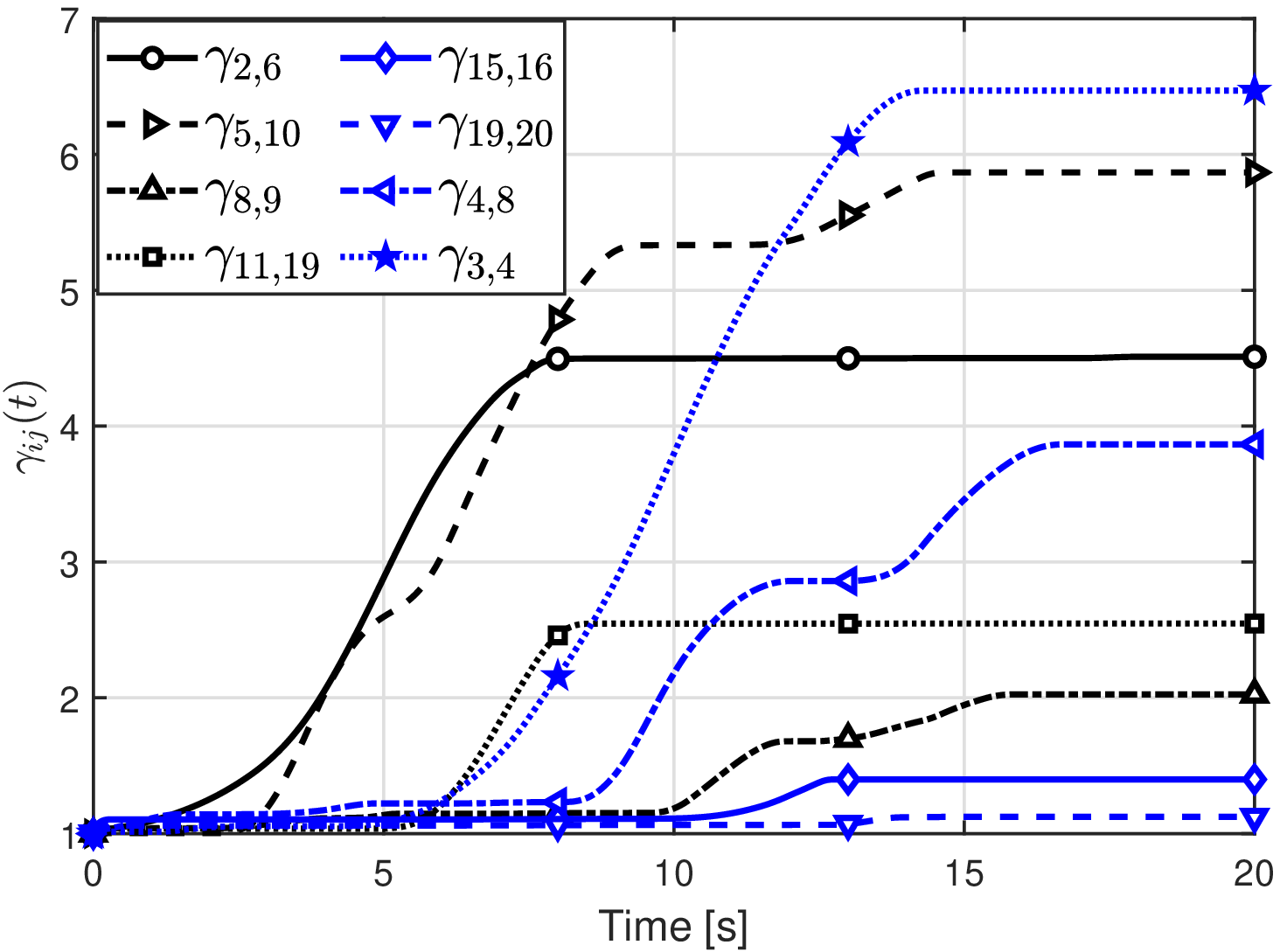}
\label{fig:sim3gamma}}
\end{subfloat}\hfill
\begin{subfloat}[]{
\includegraphics[width=0.3\textwidth]{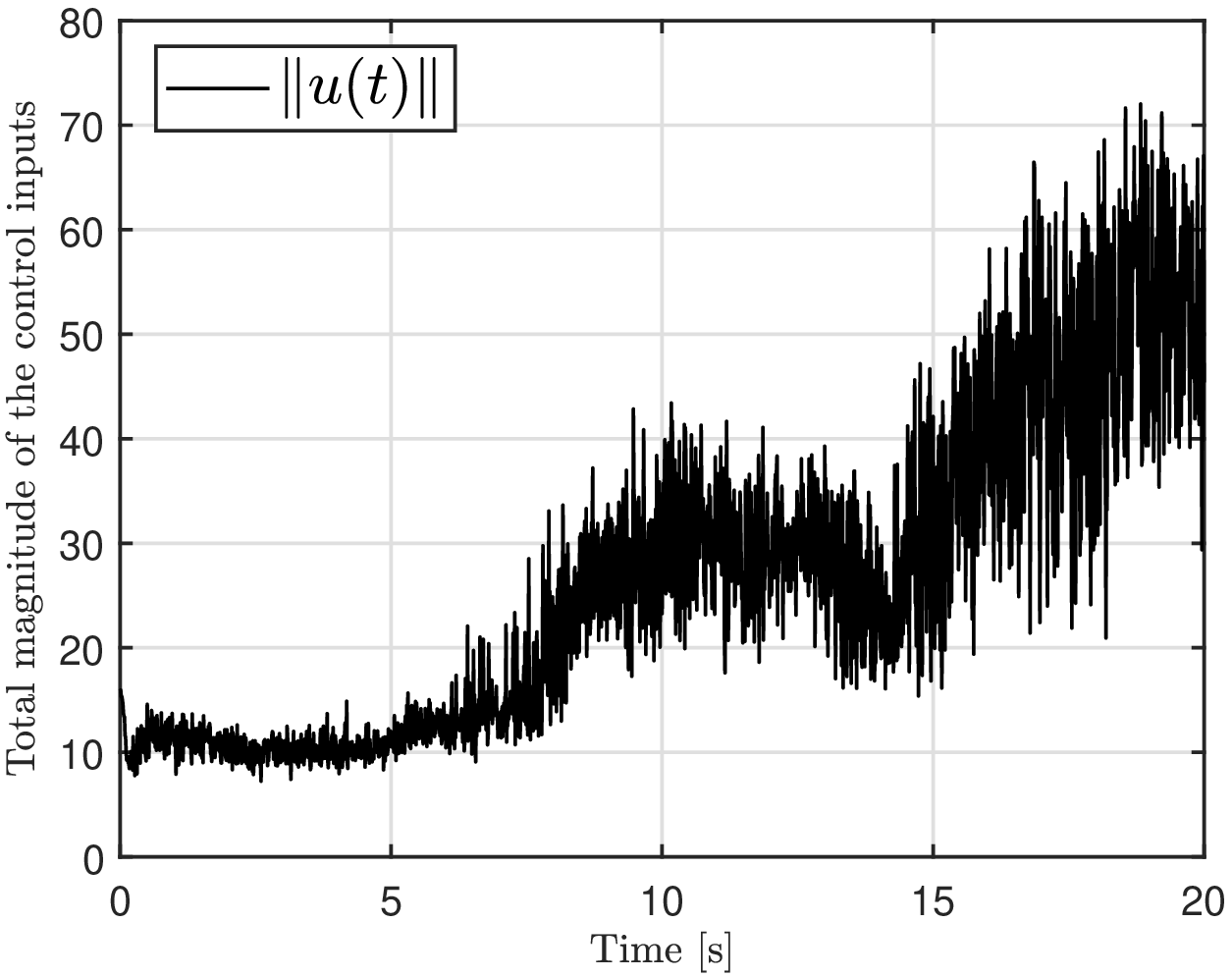}
\label{fig:sim3input}}
\end{subfloat}
\caption{Simulation 3: the 20-agent system with moving leaders under the control law \eqref{eq:bearing_based_control_law}. (a) Trajectories of agents (leaders' trajectories are colored blue, followers' positions at $t=0$ and $t=20$ sec. are marked with `${\rm x}$' and `${\rm o}$', respectively); (b)  Formation's error versus time; (c) A subset of the adaptive gains $\gamma_{ij}$ versus time; (d) The magnitude of the control input versus time. \label{fig:Sim3}}
\end{figure*}

Simulation results are shown in Fig.~\ref{fig:Sim3}. It can be seen from Fig.~\ref{fig:sim3err} that for $t\leq 6$ seconds, the formation's error increases because the adaptive control gains $\gamma_i(t)$, which specify magnitude of the control input, is still quite small. For $t\geq 6$ second, the formation's error $\boldsymbol{\delta}$ decreases to 0. Fig.~\ref{fig:sim3gamma} shows that the adaptive gains tend to increase for $0\leq t \leq 17$ second, and after the desired formation has been achieved (approximately at $t=17$ second), $\gamma_i(t)$ remain unchanged. The magnitude of the control input $\|\m{u}(t)\|$ versus time is correspondingly displayed in Fig.~\ref{fig:sim3input}, which varies accordingly to the adaptive gains and the leaders' velocity.

\subsection{Bearing-only formation tracking}
In this subsection, we simulate the formation with moving leaders \eqref{eq:bearing_only_maneuver_system2}. The initial positions of the agents and the leaders' velocities are chosen the same as the previous simulation in subsection \ref{subsec:6.3}. The simulation's parameters are $\kappa = 2$, $\gamma_i(0)=0.5$. The disturbances acting on agents $i$ are chosen as
\begin{align}
\m{d}_i(t) = \begin{cases}
\m{0}_3, & \mbox{if } 0\leq t \leq 15 s, \\
3 \m{h}_i(t), & \mbox{if } t \geq 15 s.
\end{cases}
\end{align}

\begin{figure*}
\centering
\begin{subfloat}[]{
\includegraphics[width=.98\textwidth]{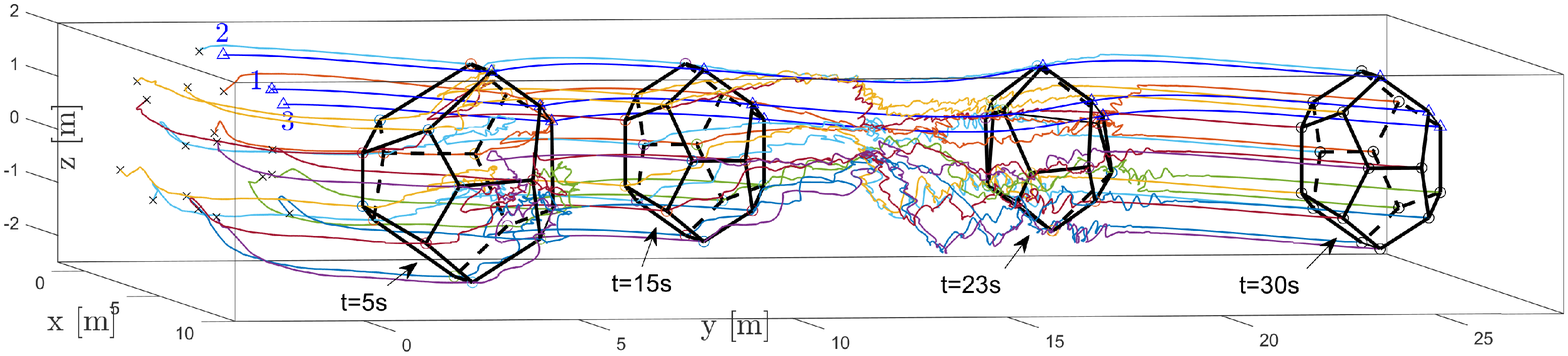}
\label{fig:sim4traj}}
\end{subfloat}\\
\begin{subfloat}[]{
\includegraphics[width=0.3\textwidth]{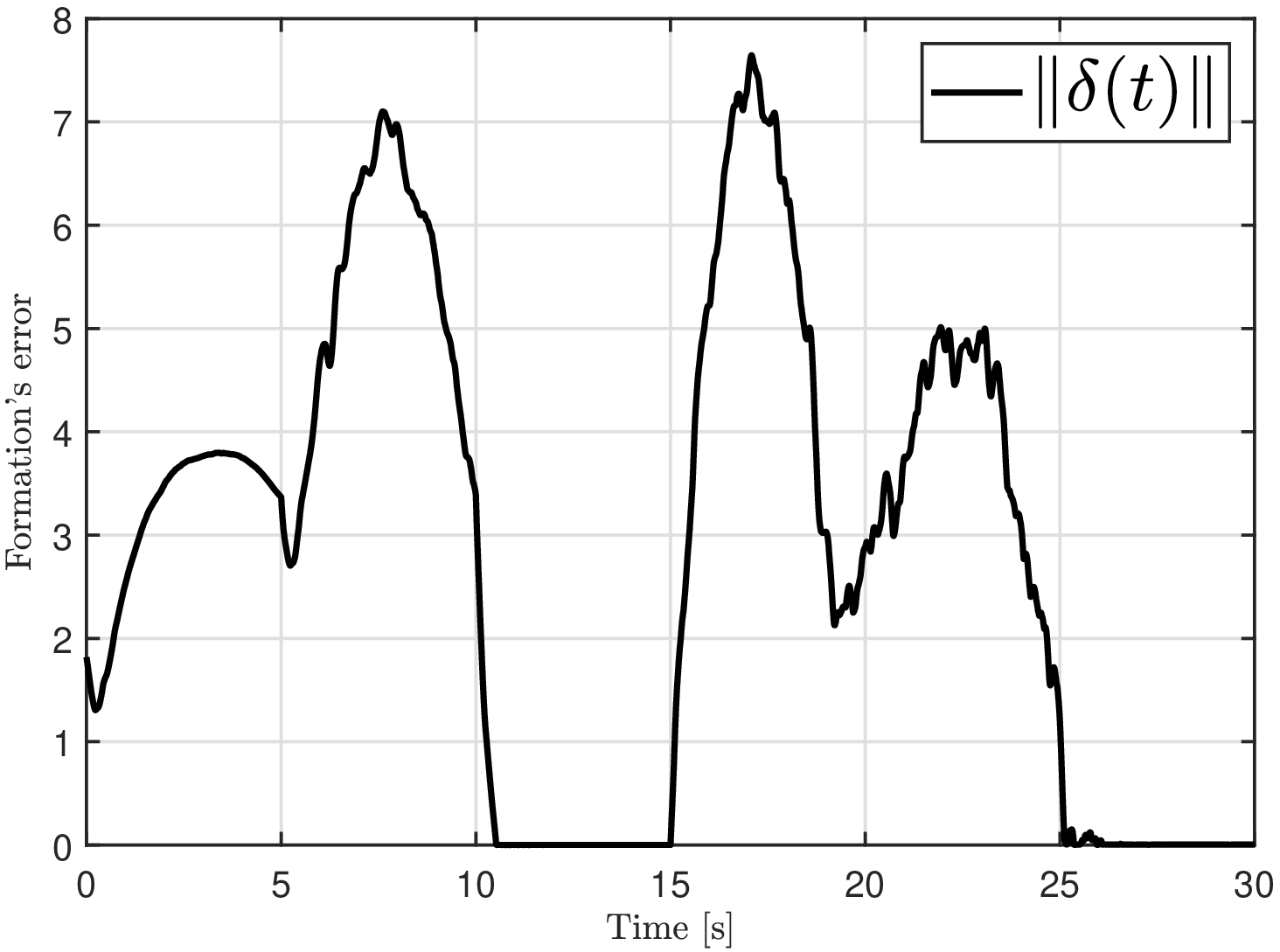}
\label{fig:sim4err}}
\end{subfloat}\hfill
\begin{subfloat}[]{
\includegraphics[width=0.3\textwidth]{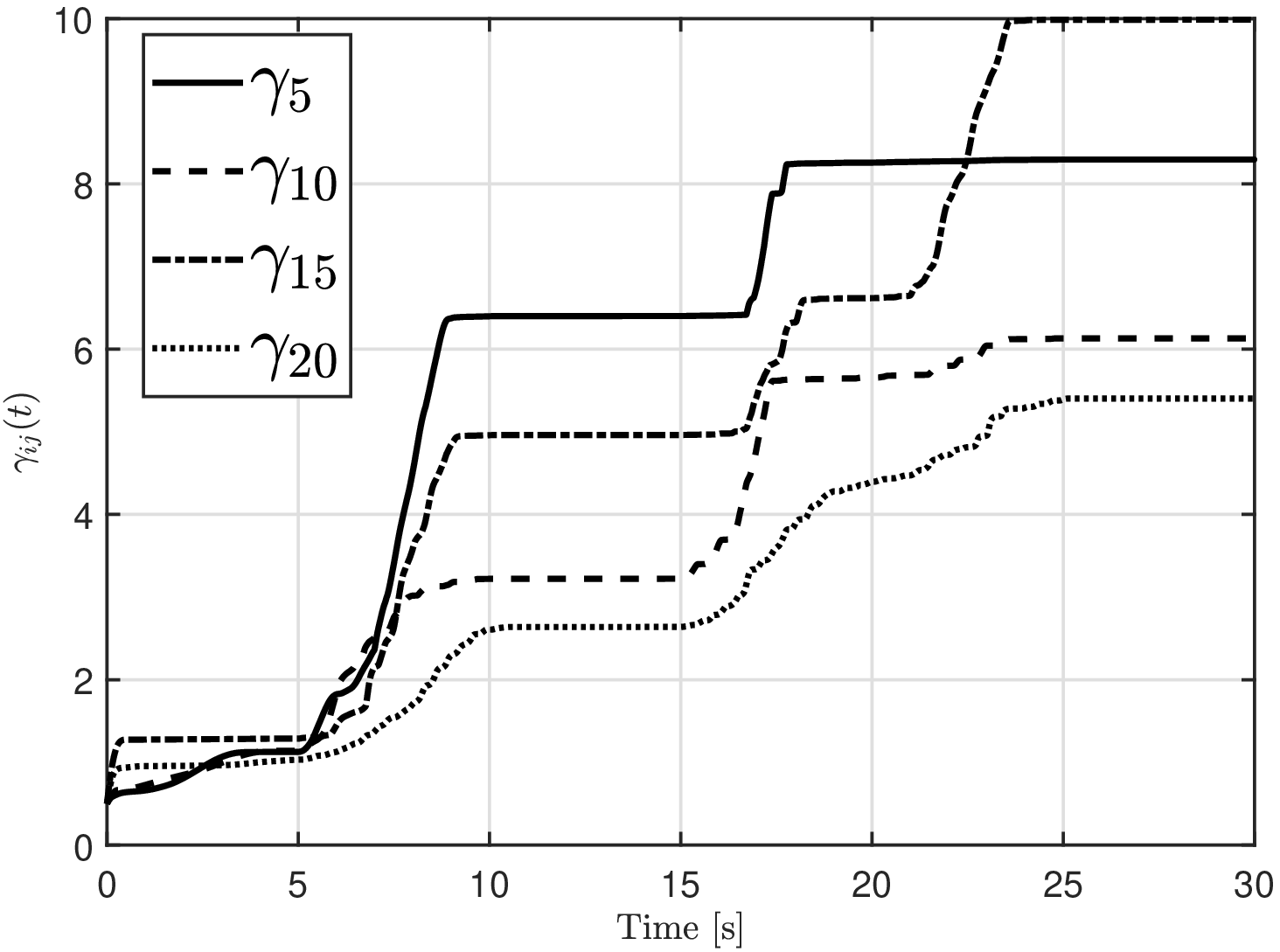}
\label{fig:sim4gamma}}
\end{subfloat}\hfill
\begin{subfloat}[]{
\includegraphics[width=0.3\textwidth]{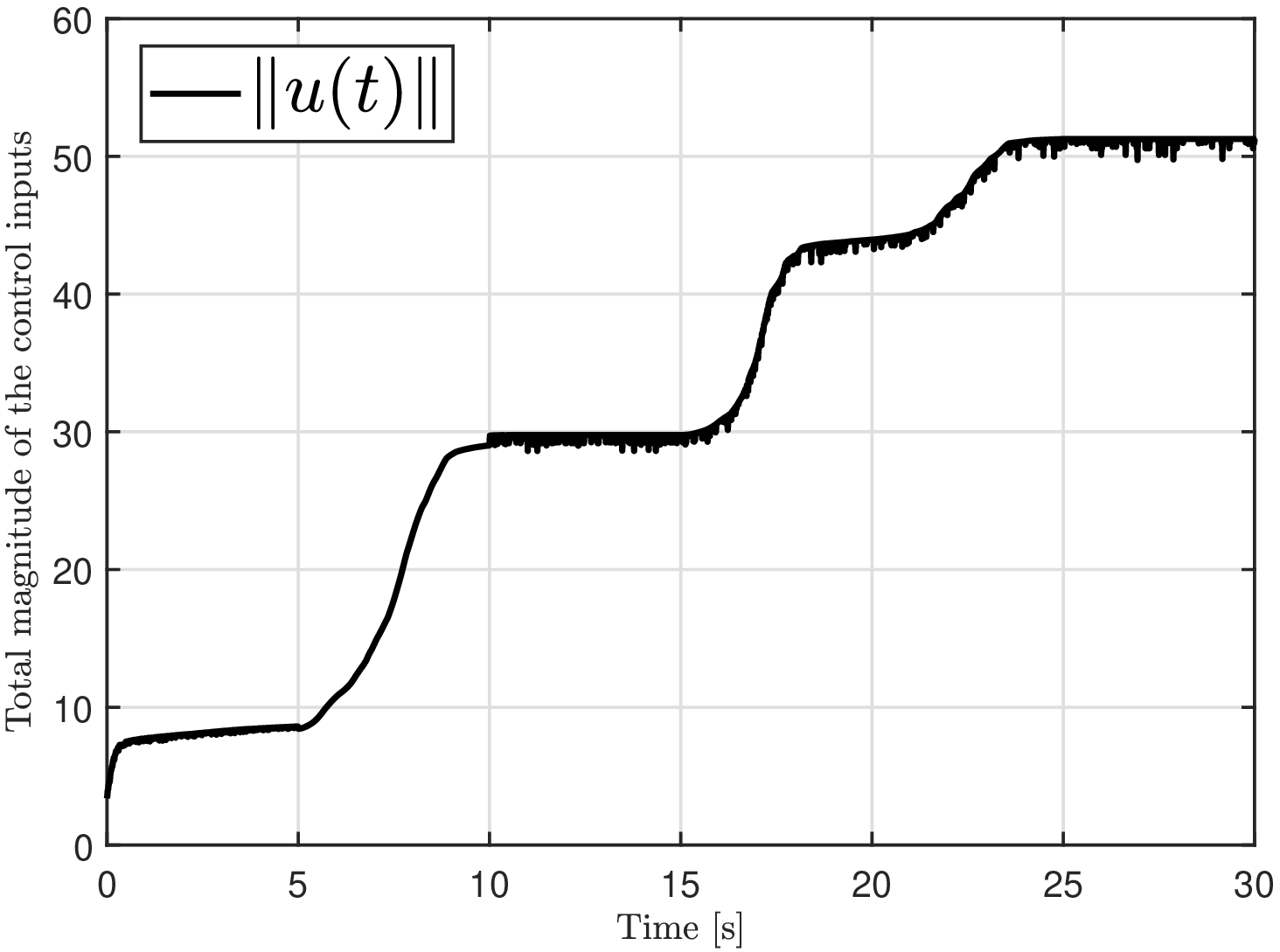}
\label{fig:sim4input}}
\end{subfloat}
\caption{Simulation 4: the 20-agent system under the control law \eqref{eq:Bearing_OnlyC} with moving leaders. (a) Trajectories of agents (leaders' trajectories are colored blue, followers' positions at $t=0$ are marked with `${\rm x}$' and at $t=5,~15,~23,~30$s are marked with `${\rm o}$', respectively); (b) Formation's error versus time; (c) A subset of the adaptive gains $\gamma_{i}$ versus time; (d) The magnitude of the control input versus time. \label{fig:Sim4}}
\end{figure*}

Simulation results are depicted in Fig.~\ref{fig:Sim4}. For $0\leq t \leq 15$s, no disturbances acting on agents, and the desired moving formation is tracked after about 11 seconds. $\gamma_i$ are increasing during this time period. The behavior of the system is quite similar to the previous simulation. However, it is 
observed that the bearing-only control law \eqref{eq:Bearing_OnlyC} 
gives a relatively faster convergence rate then the displacement-based control law \eqref{eq:bearing_based_control_law}. This can be explained by the fact that in \eqref{eq:bearing_based_control_law}, the displacement $(\m{p}_i - \m{p}_j)$ are projected into im$(\m{P}_{\m{g}_{ij}^*})$. This makes $\m{q}_{ij}$ becoming relatively small, especially when the angles between $(\m{p}_i - \m{p}_j)$ and $\m{p}_i^* - \m{p}_j^*$ are small. In contrast, the control law \eqref{eq:Bearing_OnlyC} uses only the sign of the bearing error, which is dimensionless.

For $t\geq 15$s, due to the presence of the disturbances, $\m{p}$ temporally cannot track $\m{p}^*$ (Figs. \ref{fig:sim4traj}--\ref{fig:sim4err}). Correspondingly, as depicted in Figs. \ref{fig:sim4gamma}--\ref{fig:sim4input}, the adaptive gains $\gamma_i$ and the control magnitude $\|\m{u}(t)\|$ increase again. As $\gamma_i$ is large enough, the control law simultaneously rejects the disturbance and renders the agents to their desired moving target point (approximately after 27 seconds). 

\section{Conclusions}
\label{sec:6}
The bearing-constrained formation control with unknown bounded disturbances has been studied for two types of measurements: displacements and bearing vectors. The proposed control laws can adapt the control magnitudes separately for each bearing constraint whenever the desired constraint has not been satisfied. Once the control magnitudes have exceeded the magnitude of the disturbances, it is possible to stabilize the desired configuration in finite time. Since the disturbance's magnitude may increase after the desired formation has been achieved, it may temporarily make the agents leave the desired configuration. The magnitude of the control laws will then increase accordingly to cope with the disturbances and eventually stabilize the target formation again. This process can be repeated as long as there is disturbance and control gains which always depend on the constraints' errors. Several modifications of the proposed control laws with regard to the upper bounds of the matched disturbance and the error’s bound have been also discussed. Notably, the formation tracking problem with unknown bounded leaders' velocity can be also solved with the proposed control framework. 

The theoretical results on bearing-based formation control has been rapidly filled up in recent years. Further research interests will gradually be shifted toward the implementation of bearing-only algorithms on formations of unmanned vehicles, possibly by combining the state-of-the-art theoretical findings with vision-based and machine-learning techniques.

\bibliographystyle{abbrv}
\bibliography{reference}
\end{document}